\documentclass[emulateapj,twocolumn]{aastex62}
\usepackage{graphicx}
\usepackage{float}
\usepackage{amsmath}
\usepackage{subfigure}
\usepackage{natbib}

\usepackage{booktabs}

\begin{document}

\title{Spectacular HST observations of the Coma galaxy D100 and star formation in its ram pressure stripped tail}
\author{Cramer, W. J.}
\affil{Department of Astronomy, Yale University, New Haven, CT 06511, USA}
\author{Kenney, J.D.P.}
\affil{Department of Astronomy, Yale University, New Haven, CT 06511, USA}
\author{Sun, M.}
\affil{Department of Physics, University of Alabama in Huntsville, Huntsville, AL 35899, USA}
\author{Crowl, H.}
\affil{Bennington College, Bennington, VT 05257, USA}
\author{Yagi, M.}
\affil{Optical and Infrared Astronomy Division, National Astronomical Observatory of Japan, 2-21-1, Osawa, Mitaka, Tokyo, 181-8588, Japan}
\author{J{\'a}chym, P.}
\affil{Astronomical Institute, Czech Academy of Sciences, Bo\v{c}n{\'i} II 1401, 141 00 Prague, Czech Republic}
\author{Roediger, E.}
\affil{Milne Centre for Astrophysics, Department of Physics \& Mathematics, University of Hull, Hull, HU6 7RX, UK}
\author{Waldron, W.}
\affil{Department of Physics, University of Alabama in Huntsville, Huntsville, AL 35899, USA}
%\affil{University of Chicago}
%\email{wcramer@uchicago.edu}

%\author{Britt Lundgren\altaffilmark{2}}
%\affil{University of Wisconsin -- Madison}

%\altaffiltext{1}{Advisor on this project.}
%\altaffiltext{2}{ Digital Sky Survey (SDSS)}

\begin{abstract}
We present new HST F275W, F475W, and F814W imaging of the region of the Coma cluster around D100, a spiral galaxy with a remarkably long and narrow ($60 \times 1.5$ kpc) ram pressure stripped gas tail. We find blue sources coincident with the H$\alpha$ tail, which we identify as young stars formed in the tail. We also determine they are likely to be unbound stellar complexes with sizes of $\sim$ $50 - 100$ pc, likely to disperse as they age. From a comparison of the colors and magnitudes of the young stellar complexes with simple stellar population models, we find ages ranging from $\sim$ $1 - 50$ Myr, and masses ranging from $10^3$ to $\sim$ $10^5$ M$_{\odot}$. We find the overall rate and efficiency of star formation are low, $\sim$ $6.0 \times \, 10^{-3}$ $M_{\odot}$ yr$^{-1}$ and $\sim$ $6 \, \times$ 10$^{-12}$ yr$^{-1}$ respectively. The total H$\alpha$ flux of the tail would correspond to a SFR $7$ times higher, indicating some other mechanism for H$\alpha$ excitation is dominant. From analysis of colors, we track the progression of outside-in star formation quenching in the main body of D100, as well as its apparent companion the S0 D99. Finally, we observe the dust extinction in the base of the tail has an outer envelope with remarkably smooth and straight edges, and linear filamentary substructure strongly suggestive of magnetic fields. These features and the small amount of tail broadening strongly suggest gas cooling restricting broadening, and the influence of magnetic fields inhibiting turbulence.
\end{abstract}

\section{Introduction}

The recent discovery of a number of remarkable tails of gas and stars stripped from galaxies found in clusters has been an exciting area of study in the field of cluster galaxy evolution e.g. \citet{Gavazzi+95, Kenney+99, Gavazzi+01, Yoshida+02, Yoshida+04, Sun+05, Oosterloo+05, Cortese+06, Chung+07, Cortese+07, Sun+07, Kenney+08, Yoshida+08, Smith+10, Yagi+10, Abramson+11, Yoshida+12, Fossati+12, Yagi+13, Jachym+14, Kenney+14, Ebeling+14, Yagi+15, Poggianti+17, Yagi+17, Boselli+18}. Ram pressure stripping (RPS) is an important mechanism by which galaxies, especially those in clusters, evolve. It removes gas, quenches star formation, and drives evolution in environments with a sufficient density of intracluster gas \citep{Gunn+72}. In sufficiently high-mass clusters, such as Coma ($\sim$ 10$^{15}$ M$_{\odot}$), star forming galaxies can eventually be stripped of even the most strongly gravitationally bound gas at the center, completely quenching star formation \citep{BravoAlfaro+00, Smith+10, Yagi+10}.

While many recently observed ``jellyfish galaxies" have tails seen in ionized gas \citep{Yoshida+12, Yagi+10, Zhang+13, Poggianti+17}, at least some of these tails are composed of a multiphase mixture of gas, some hot enough to see in the X-ray \citep{Sun+05, Sun+10, Zhang+13, Sanders+13, Sanders+14}, but also a significant fraction by mass of molecular gas \citep{Jachym+14, Jachym+17}. Simple physics as well as simulations strongly suggest that ram pressure stripping in clusters is too weak to strip the densest gas directly, except possibly in extreme cases; thus, most of this molecular gas likely cools $\textit{in situ}$ \citep{Tonnesen+09, Tonnesen+10}. Gas cooling, heating, compression, and possibly other factors affect the locations and rates of star formation in tails, and the combination of factors is not well understood \citep{Tonnesen+12}. A number of gas-stripped tails with ongoing star formation within the tails have been discovered in massive clusters such as Coma, however, stripped tails with active star formation have also been found in lower mass clusters such as Virgo \citep{Hester+10, Fumagalli+11, Yagi+13, Kenney+14}. However, not all strongly stripped galaxies are `jellyfish' i.e. have lots of star formation in the tail (e.g. \citet{Boselli+16}).

In depth studies of RPS tails to constrain the quantity of stripped gas, as well as the mass of stars formed that eventually become part of the intra-cluster light (ICL), or fall back into the galaxy to form new thick disk or halo components \citep{Abramson+11}, are key to gaining a fuller picture of star formation in stripped tails. However, the rate and efficiency of star formation in these tails is still a matter of discussion; most findings so far point to a significantly lower efficiency of star formation in extraplanar regions than the disk of the galaxy. In a sample of Virgo galaxies studied by \citet{Vollmer+12} the star formation efficiency (SFE) was found to be $\sim$ 3 times lower in the stripped extraplanar gas than the disk. A similar study using GALEX FUV and H I maps of eight RPS tails in the Virgo cluster found the overall SFE to be $\sim$ 10 times lower in the tail than in the body of the host galaxy \citep{Boissier+12}. In \citet{Jachym+14} the SFR (SFR) surface density varied by a factor of $\sim$ 50 along three pointings of the ram pressure stripped tail of ESO 137-001 in Abell 3627, becoming less efficient with distance from the host galaxy. \citet{Jachym+14} and \citet{Vollmer+12} both used H$\alpha$ emission as a direct proxy for star formation; \citet{Vollmer+12} did also compare with SFR derived from FUV observations. The connection between H$\alpha$ emission and recent star formation in the disks of galaxies is well known, but in the unique environment of ram pressure stripped tails, subject to heating from shocks and other mechanisms, we cannot assume the same relationship holds. MUSE studies, such as that conducted by \citet{Fossati+16} showed that most of the H$\alpha$ in the tail of the galaxy ESO 137-001 was not associated with H II regions or star formation. Furthermore, in \citet{Boselli+16}, the authors concluded that the long, H$\alpha$ tail of NGC 4569 in the Virgo cluster contained no H II regions or evidence of star formation. This implies that some other ionization mechanism must be at play. It should be noted that some tails have shown good concordance between expected H$\alpha$ luminosity and SFR as measured by identifying H II regions with line ratios, such as in the jellyfish galaxy JO206 in the IIZW108 cluster \citep{Poggianti+17}. However, even if line ratios are measured and show that H$\alpha$ is the product of star formation, the H$\alpha$ may yield an underestimate of the true SFR. H II regions in RPS tails may be stripped of some of the gas close to the ionizing stars, and as a result leak more Lyman continuum photons than typical disk H II regions (Kenney, J.D.P., et al. \textit{in prep}). In sum, it is risky to derive SFRs solely through H$\alpha$ luminosities, and/or line ratios.

The ideal way to study the ages and properties of stars in RPS tails is through direct observations of the stars themselves. With an accurate measurement of the ages and locations of young stars throughout the tail, along with an estimate of the age of the host RPS tail, one can investigate the conditions leading to star formation in the tail interstellar medium (ISM). For example, in \citet{Tonnesen+12}, the authors found that in a simulation of star formation in an RPS tail, it took $\sim$ 200 Myr for tail gas to cool sufficiently for star formation to take place. However, many RPS tails show star formation occurring near the body of the galaxy, a region that contains recently stripped gas (see Figure 2 of \citet{Jachym+14} for an example), as well as no clear trend in the age of stars formed with distance along the tail (e.g. \citet{Cortese+07}). The disparity between simulation and observation suggests heating and cooling mechanisms in tails are still poorly constrained. 

Careful investigation of the location and history of star formation in tails is key to solving the disparity between simulations and observations. Studies based on direct observation of star clumps in tails have been conducted, and have found masses ranging from $\sim 10^3-10^6$ M$_{\odot}$ and ages between $\sim$ $1-100$ Myr \citep{Cortese+07, Yagi+13, Boselli+18}. Some studies have claimed older populations, although the amount of dust extinction in tails, uncertainty in the stellar formation history (burst vs. continuous model such as in \citet{Yoshida+08}), and the contamination of possible tail sources with background galaxies in the tail direction can also complicate studies of star formation in tails. For example, \citet{Fumagalli+11} estimated the ages of stars in the tail of IC3418 in the Virgo cluster to be up to 1 Gyr. However, it was shown in \citep{Kenney+14} that some of these measurements were likely overestimates due to contamination from background galaxies, and the true maximum age of stars may be closer to $\sim$ 300 Myr.

Especially important for estimating the ages of stars in tails is a UV filter, to capture the young stars, and differentiate from background sources in the cluster. With our deep HST observations in multiple bands, including the F275W UV filter, we can estimate the ages and masses of young stars in the tails. We also have the resolving power, and depth, to differentiate background galaxies from tail sources, and analyze dust extinction in the most recently stripped part of the tail. We also make use of the resolution of HST to measure the sizes of stellar associations to estimate whether they are gravitationally bound. If they are not gravitationally bound, star clumps will dissociate over time, falling below detection limits, and mixing with the ICL. This possibility has not been studied previously. The fantastic resolution of HST allows us to do so.

\begin{figure*}
	\plotone{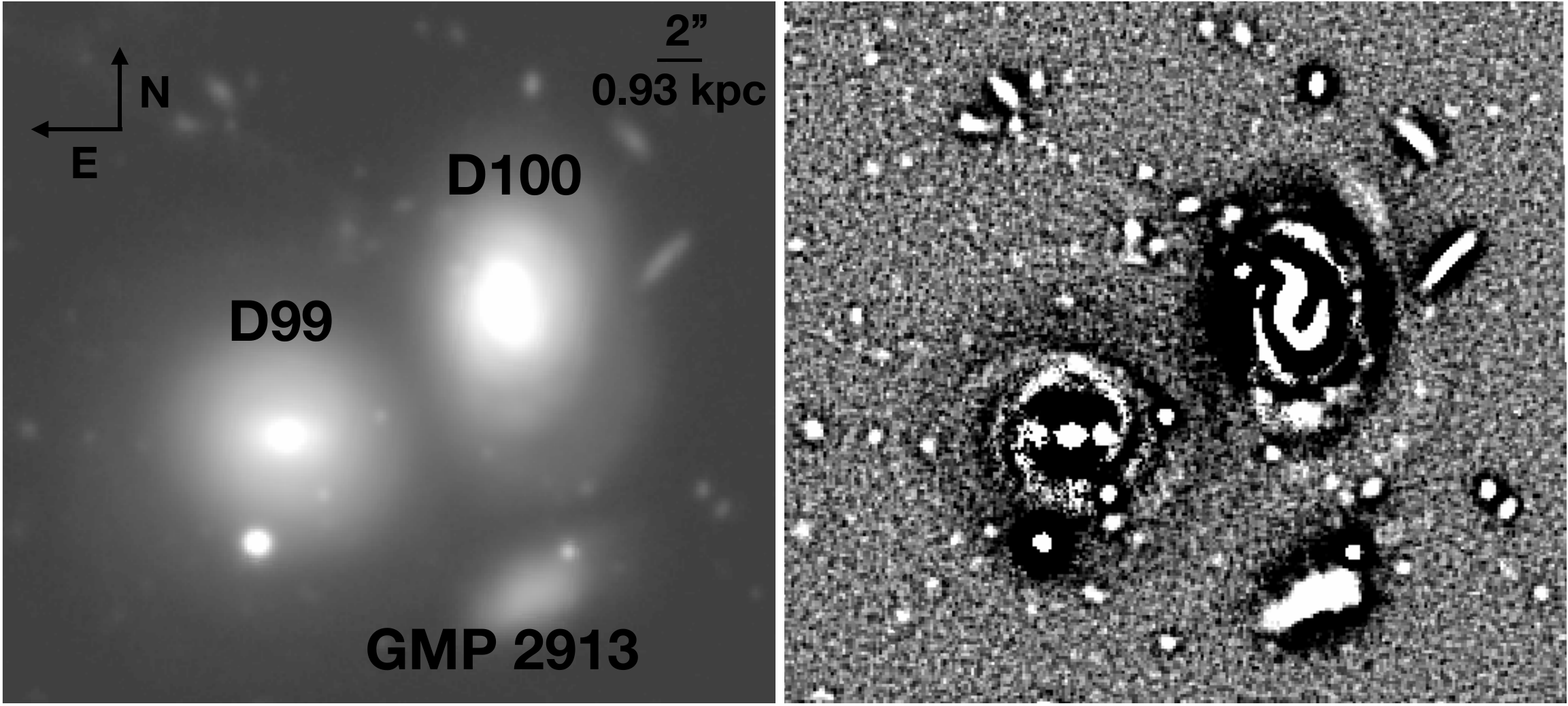}
	\caption{On the left an image of D100 and its companion galaxies, observed in the Subaru telescope $R$-band. On the right is an unsharp mask of the $R$-band image on the left, to highlight substructure.}
	\label{fig:Subaru}
\end{figure*}

\subsection{D100}

D100, named in \citet{Dressler+80} (also identified as GMP 2910 \citep{Godwin+83}), is a SBab galaxy in the Coma cluster, with a luminosity of 0.3 L$_{\star}$, and an estimated stellar mass of 4 $\times \, 10^9$ M$_\odot$ from the WiscM11 spectral energy distribution (SED) model \citep{WiscM11}\footnote{The mass estimate in the catalogue had to be adjusted for the true distance to Coma, as opposed to that estimated from simply the redshift of D100, which is affected by the intrinsic velocity of the galaxy in the cluster.}. An $R$-band image from the Subaru telescope \citep{Yagi+07} of the galaxy and its neighbors is shown in Figure \ref{fig:Subaru}. The galaxy is at a projected distance of $\sim$ 240 kpc from the center of the Coma cluster. We adopt a luminosity distance to Coma of 100 Mpc, corresponding to a size scale of 1''=0.464 kpc, and a distance modulus of 35.0, using standard cosmological parameters from the WMAP nine year survey \citep{Hinshaw+13}. D100 has a remarkable 60 kpc long and extremely narrow, 1.5 kpc wide, tail of gas that streams out from the center of the galaxy, first observed in H$\alpha$ by the Subaru Suprime-Cam \citep{Yagi+07}, and seen in Figure \ref{fig:HST_Ha}. This galaxy was also observed with the HST WFC2 by \citet{Caldwell+99}, who noted strong ongoing star formation, observed with ground-based spectroscopy, in the central 2'' ($\sim$ 1 kpc) of the galaxy, but a $\sim$ 0.25 Gyr post-starburst spectrum at a radius of $\sim$ 3''. We note that \citet{Caldwell+99} did not have HST UV data, and their optical HST data was shallower and with poorer resolution than our new WFC3 data. The authors pointed out the ``unusual morphology" of the prominent dust on the northern side of D100, perhaps a hint of the tail that would be discovered in the later H$\alpha$ observations by \citet{Yagi+07}. The tail and body of the galaxy were observed in the UV with GALEX and CFHT MegaCam by \citet{Smith+10}. They find faint $u$-band emission along the length of the tail with some more concentrated knots, suggesting young stars. However, the 3727\AA [OII] doublet falls in the CFHT $u$-band making it difficult to establish the direct correlation of $u$-band emission to the presence of young stars, due the possible presence of warm gas in the spectrum. Deeper optical imaging of the tail is necessary to investigate the presence of stars.

The components of the gaseous tail of D100, as well as some properties of the tail that resulted from ram pressure were investigated in detail in \citet{Jachym+17}. Along with hot gas detected in the soft X-ray, large amounts of molecular gas were found along the length of the tail identified via observations with the IRAM 30m telescope in CO(2-1) and CO(1-0) in four different pointings with a 30'' beam. The total mass of molecular gas in H$_2$ in the tail is estimated to be $\sim$ 1 $\times$ 10$^9$ M$_\odot$, assuming a standard value for the relation between CO emission and H$_2$ mass. The tail was undetected in HI to a 3$\sigma$ limit of $\sim 0.5 \times 10^8$ M$_\odot$, suggesting that the tail is dominated by cold molecular gas. \citet{Jachym+17} note that while they find abundant molecular gas, they are only able to state that star formation in the tail \textit{may} be present, due to the inability to distinguish between H$\alpha$ arising from photoionization by hot stars, and that excited by some other method.

\subsection{Outline}

In Section 2 of this paper, we present the details of our HST observations and the data reduction scheme employed. In Section 3 we describe the HST images of the three galaxies in the field, noting features and evidence for ram pressure stripping, as well as outside-in quenching. We analyze the morphology of the RPS tail of the galaxy through investigation of the dust in the tail. We also constrain the SFR and SFE of the tail, as well as the characteristics of the star clumps in the tail. Finally, in Section 4, we discuss the astrophysical effects contributing to the properties of the RPS tail catalogued in Section 3. In Section 5 we summarize our results.

%\keywords{QSOALS, cosmic metallicity, SN: type Ia}

\section{Observations}

D100 was observed as part of our HST program 14361 (PI: Sun), targetting the region near the center of the Coma cluster with the HST ACS instrument in F475W for $1440$ seconds and F814W for 674 seconds, as well as the WFC3 instrument in F275W for 2583 seconds, in May-July of 2016. The image was centered such that we were able to see the full extent of the H$\alpha$ tail identified in \citet{Yagi+07}. Part of the same field of view, including the body of D100 and its tail, was covered in parallel observations in late 2017 and early 2018 in HST program 14182 (PI: Puzia), adding 2396s of F814W time, and 4454s of F475W time. While D99 and GMP2913 were covered only by our original data, D100 and the tail were covered for a total of 5894s in F475W and 3070s in F814W. The ACS data in two high-throughput filters allows us to detect faint optical features, as well as make a detailed color map of the galaxy. The F275W filter allows us to trace the light from young stars in order to identify recent star formation in the body and tail of the galaxy. When combined, we can use the three filters to constrain the stellar ages and masses of star clumps found in the tail, as well as the star formation history of the disk. The H$\alpha$ data we used have been corrected for the velocity gradient of the tail, over-subtraction in the $R$-band, and for the presence of [NII] and [SII] lines in the filter, with the assumption that [NII]/H$\alpha$=0.66, and [SII](6717\AA+6731\AA)/H$\alpha$=0.66. The correction method was the same as that described in \citet{Yagi+17}.

Our observations reached a limiting surface brightness, estimated by taking the value of three times the background standard deviation, of 29.1 mags in F275W, 30.0 mags in F814W, and 30.9 mags in F475W. Magnitudes listed throughout this paper are given in AB mags. Milky Way extinction is minimal in the direction of Coma; from \citet{Schlafly+11} estimates for SDSS filters that roughly  correlate to ours, extinction in the $u$-band, or F275W, is 0.035 mags, in $g$-band, or F475W, 0.027 mags, and in the $r$-band, F814W, 0.019 mags.

\subsection{Data Reduction}

To combine the data from our original proposal with the new data from the parallel observations we used both the HST pipeline tools with DrizzlePac, and the astrodrizzle tool SWarp \citep{SWarp}. We first aligned the data from the different fields of view with DrizzlePac, as well as removing some cosmic rays from fields covered by multiple observations. After alignment, the images were drizzled, a process standard to HST imaging by which images are combined while preserving photometry and resolution, weighted by the statistical significance of each pixel, and corrected for geometric distortions. The data was also corrected for charge transfer effects using DrizzlePac, then combined using SWarp. Because the F275W data had a smaller pixel scale than the F475W and F814W observations, the F475W and F814W had to be re-gridded to the pixel scale of the F275W so that the images could be compared on a pixel-pixel basis. This was done using the Lanczos-3 6x6-tap filter interpolation scheme in SWarp, the recommended method for flux-conserving interpolation.

In areas not covered by multiple observations, such as the regions around D99 and GMP 2913, obvious cosmic rays were still visible in the image, even after the automatic HST pipeline was run. As only two pointings were taken in each filter, and each was slightly offset from the other, there is a detector gap in the WFC3 and ACS imagers that is only covered once. As such, the standard HST pipeline to remove cosmic rays could not be used here, and a custom scheme needed to be employed. Sources were identified with the Astromatic tool SExtractor \citep{Sextractor+96} in each filter, and then compared between the filters. If a source was found in only one filter to higher than 12$\sigma$ significance (a typical threshold for the peak of a cosmic ray in HST data) it was labeled a cosmic ray and cleaned from the image. If the source was detected in more than one filter it was not removed.

\begin{figure*}
	\plotone{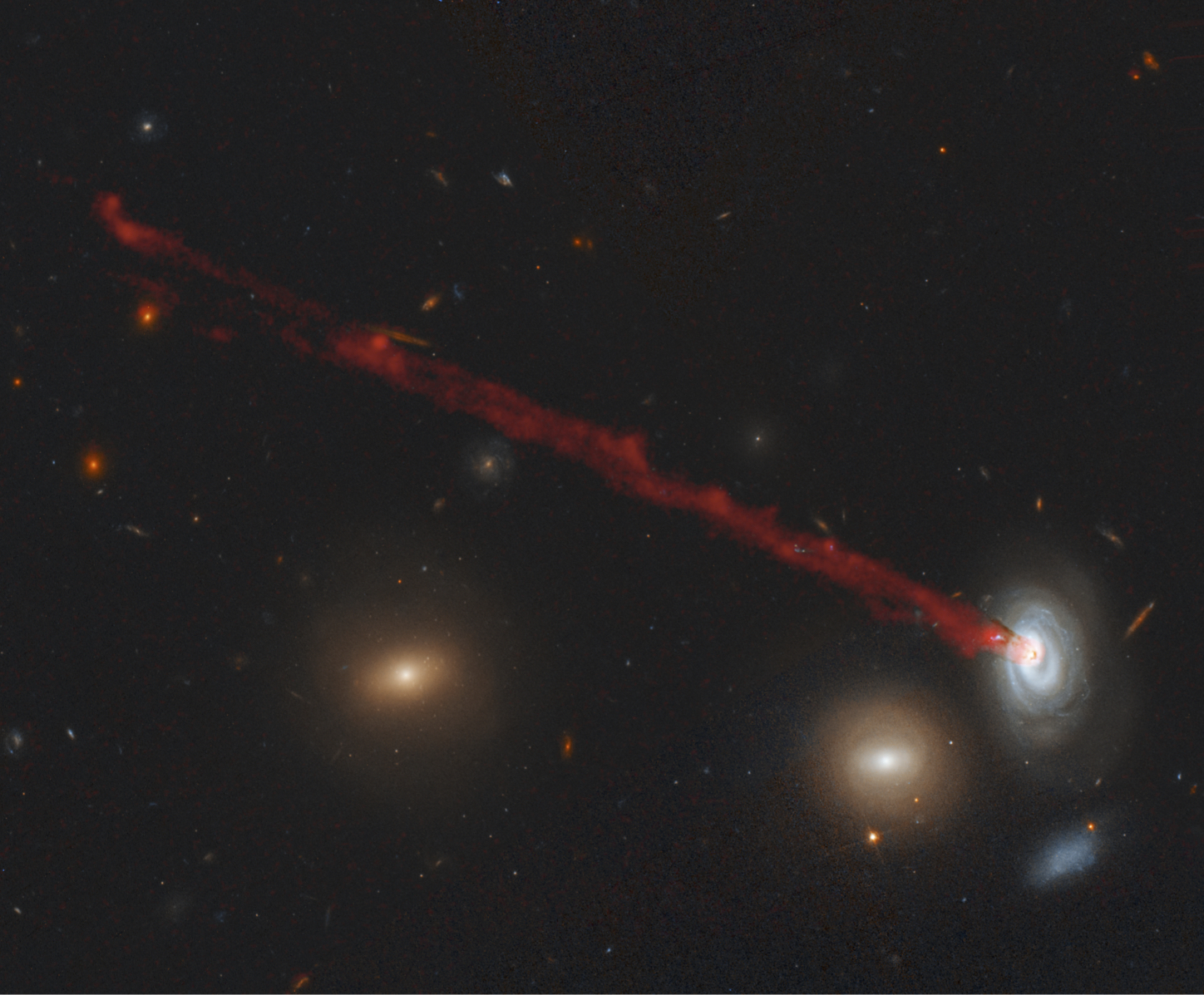}
	\caption{An HST false color image with the F814W filter in red, the F475W filter in blue, and an average of the F475W and F814W in green. The F275W was not used for this image, as the signal to noise in the F275W was much lower than the other filters. Overlayed in bright red is H$\alpha$ data from the ground based Subaru Suprime-Cam, first published in \citet{Yagi+07}. The image was provided to us by the STScI imaging team led by J. DePasquale.}
	\label{fig:HST_Ha}
\end{figure*}

The resulting reduced images are shown in Figures \ref{fig:HST_Ha} and \ref{fig:sidebyside}.

%Replace with a version without the circles, possibly a lot more zoomed in

\section{Analysis}

\subsection{Main Body of D100}

\begin{figure*}
	\plotone{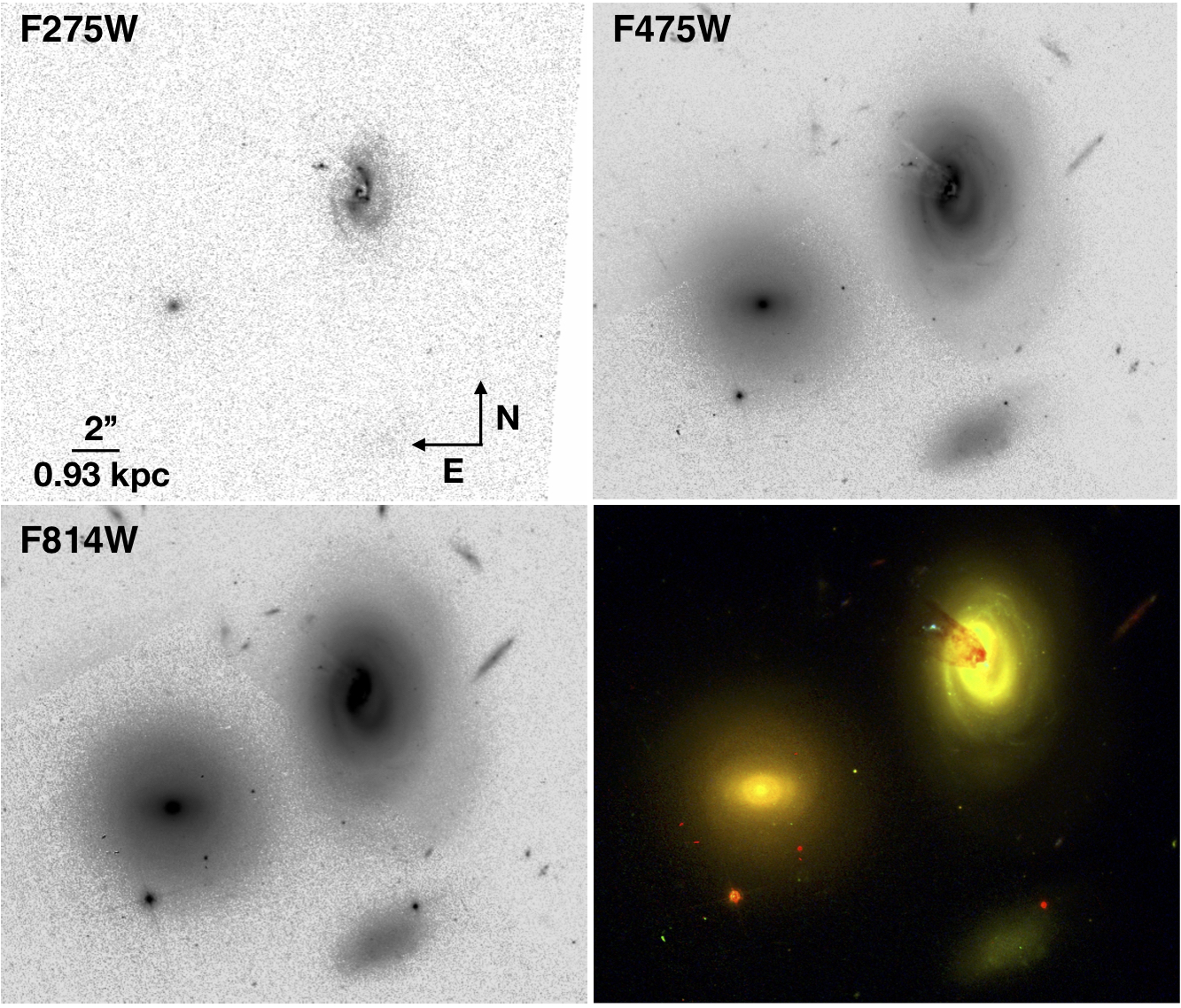}
	\caption{The images, in the three HST bands, of D100, its neighboring anemic spiral galaxy to the left, D99, and the dwarf galaxy GMP 2913 to the south. At the bottom-right is a false color image made from combining F275W (blue), F475W (green), and F814W (red).}
	\label{fig:sidebyside}
\end{figure*}

\begin{figure*}
	\plotone{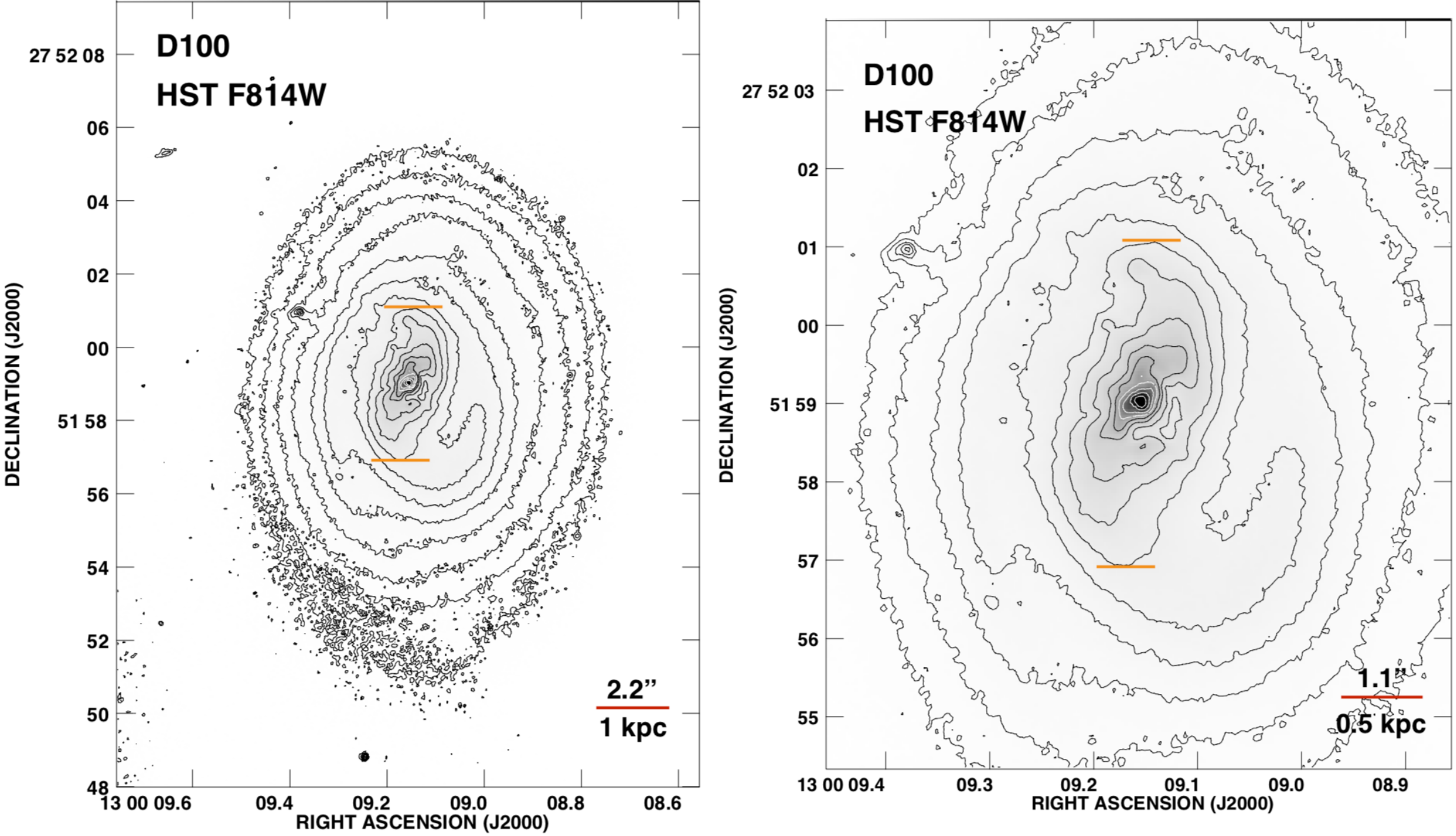}
	\caption{An isophotal contour plot of the F814W filter image of D100, shown on the left, and zoomed in on the central region on the right. The contour levels vary from 22.0 to 16.9 mag/arcsec$^2$ by increments of 0.37 mag/arcsec$^2$. The position angle of the isophotes appears to be relatively fixed in the inner part of the galaxy, bounded by the two orange marks, suggesting a bar-like structure. Beyond this, the isophotes follow the structure of the spiral arms out a few contours, then eventually become more consistently elliptical.}
	\label{fig:isophote}
\end{figure*}

D100 appears to show a classical grand design spiral structure, with two prominent arms. Through isophotal analysis of the F814W data, shown in Figure \ref{fig:isophote}, we have found evidence for a bar-like structure in D100, stretching across a diameter of about 2.6'', or 1.2 kpc. The central region of the galaxy appears relatively undisturbed. It is unlikely that D100 is interacting with its neighbor to the east, D99, as a large difference in velocity (4500 km s$^{-1}$ greater than that of D100) makes interaction unlikely. However, it appears likely that D100 has had some interaction with the dwarf galaxy GMP 2913 to the south. This galaxy is roughly three magnitudes fainter than D100 and D99 in the F814W. At a distance of 100 Mpc, we measure its absolute magnitude as $-16.65$ in F814W, and it has a velocity difference with D100 of only 132 km s$^{-1}$, based on its redshift from \citet{Yagi+07}. The two galaxies show evidence of interaction, as the southwest outskirts of the disk of D100 show a clear disturbance in the deep Subaru $R$-band image (Figure \ref{fig:Subaru}) and the HST image (Figure \ref{fig:sidebyside}), as well as irregular isophotes in the HST F814W data (Figure \ref{fig:isophote}). The tidal disturbance appears to be a weak interaction; the dwarf morphology is a bit irregular, and the disruption in D100 appears limited to an elongation of the southern edge of the galaxy at $r \sim$ 6'' $\sim$ 3 kpc. Simulations from \citet{Toomre+72} showing the early stages of tidal interaction being limited to slight elongation of one side of the galaxy suggest that the interaction between D100 and GMP2913 may have begun only recently. Thus, we do not believe the tidal interaction should have had a direct impact on the ram pressure stripping of the galaxy. It is possible, however, it may have altered the distribution of gas in D100 (if this gas had not already been stripped), driving some gas inward, and making this concentrated gas harder to strip.

Information on the star formation history of the galaxy can be gleaned from simply inspecting the high-resolution HST three color image, shown in Figure \ref{fig:colorstreams}. The outer disk appears to have been stripped of all its ISM by the ram pressure, as the only clear dust extinction in D100 is visible in the the central $r \sim 1$''.  This is also where the F275W emission appears most concentrated in several pockets within $r$ $\sim$1'' of the nucleus, along the spiral arms. However, the outer parts of these arms have a redder color, indicating a stellar population older than the innermost parts of the spiral arms, indicative of outside in quenching of star formation (see Section 3.2 for further discussion). Other than this central region, the only other noticeable star formation in the galaxy is visible in the region 3'' northeast of the nucleus, in the dust tail. This bright F275W source (see Figure \ref{fig:colorstreams}) near the outskirts of the galaxy appears to be a very blue young stellar complex within the dust tail.

\subsection{Snaky stellar streams}

\begin{figure}
	\plotone{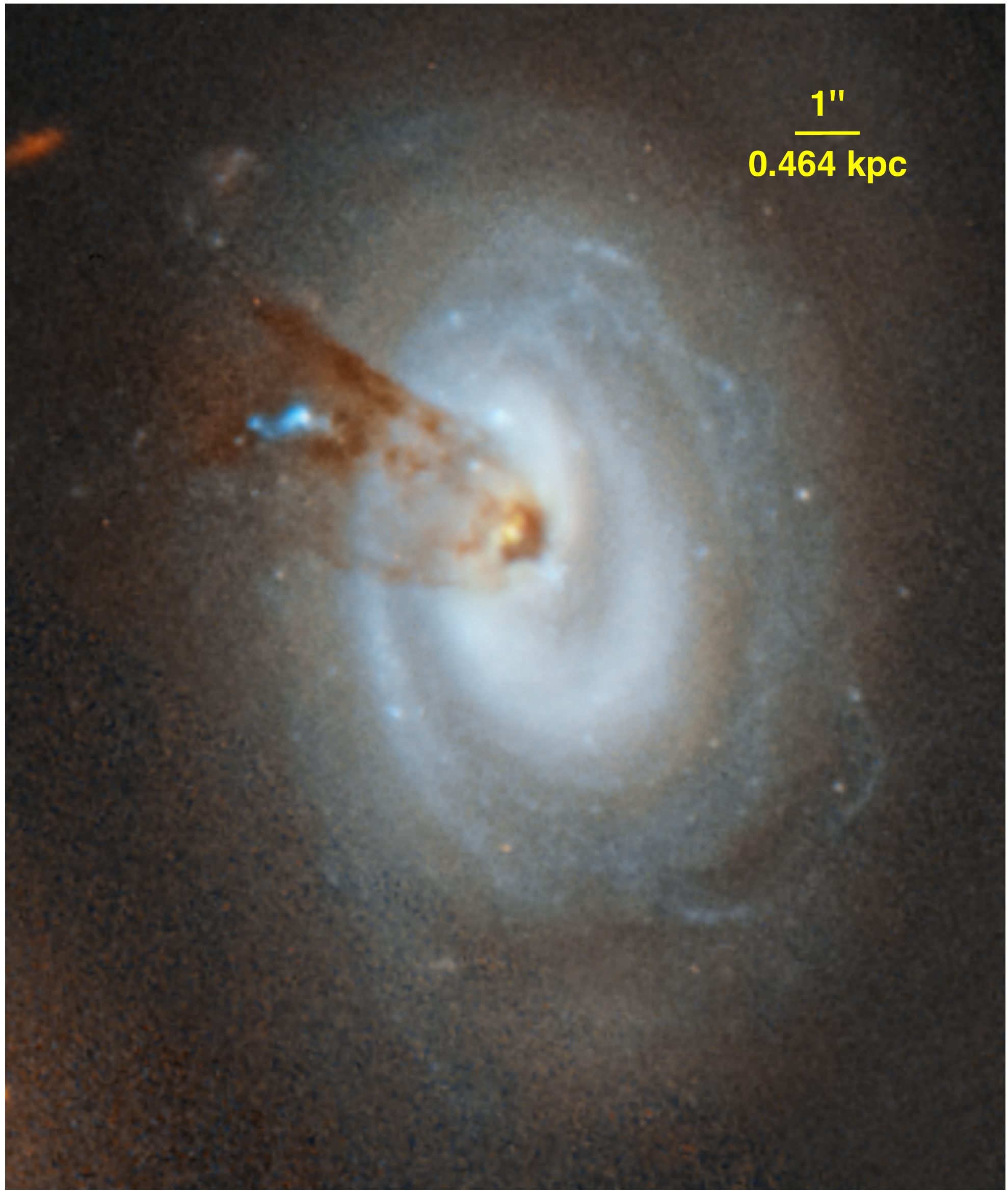}
	\caption{A false color HST image with the three filters from our study, F275W in dark blue, F475W in lighter blue, and F814W in red. The most obvious region of extraplanar star formation in the tail (Source 1) is 3'' northeast of the galaxy nucleus. Image provided by STScI imaging team led by J. DePasquale.}
	\label{fig:colorstreams}
\end{figure}

\begin{figure}
	\plotone{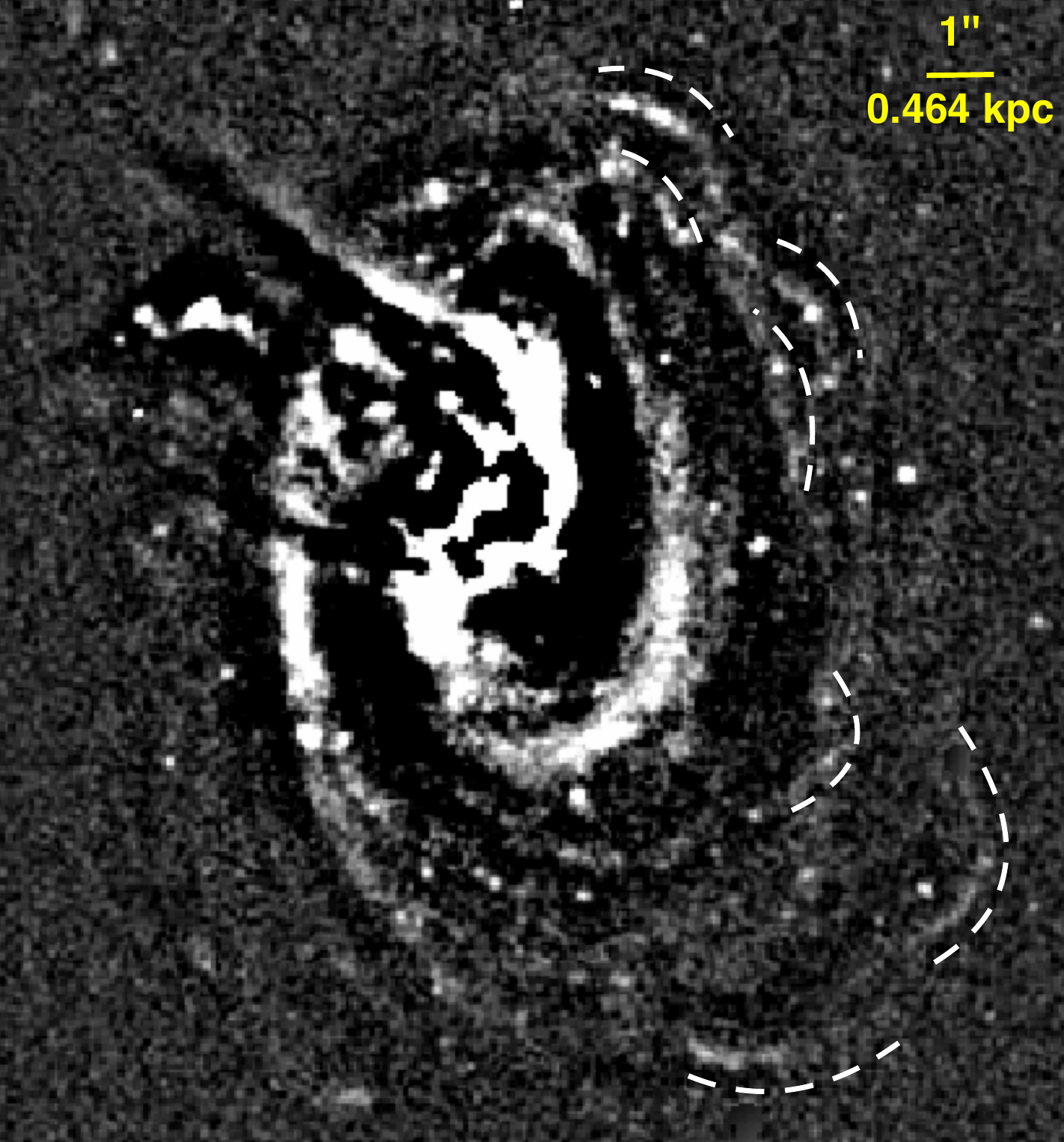}
	\caption{An unsharp-mask HST image of D100, in the region shown in Figure \ref{fig:colorstreams}, with the F475W filter image. Marked in white are thin stellar streams which appear distinct from the normal spiral structure, and which could be the result of ram pressure.}
	\label{fig:unsharp}
\end{figure}

An interesting feature in the outskirts of the galaxy, faintly visible in the color image (Figure \ref{fig:colorstreams}) are the deviations in the stellar morphology from the presiding grand design spiral structure, located at $r$ $\sim$ 4.5'' from the nucleus in the north, and $r$ $\sim$ 6'' in the southeast. Thin (width of $\sim$ 0.25'') and long (length of $\sim$ 2.5'') obliquely curved distributions of stars are visible at the outskirts of the galaxy. They are more visible in the green than the red (F475W-F814W=0.5), suggesting they are relatively young distributions when compared with the surrounding disk (F475W-F814W=1.0). In an attempt to render these thin streams of stars more visible, we also generated an unsharp mask image of the galaxy (Figure \ref{fig:unsharp}). The origin of these streams is as of yet unknown, as well as whether their formation can be related to ram pressure stripping, or tidal effects. One possibility is that they may be an abundance of stars formed in the ISM compressed by the galaxy rotating into the ram pressure front, such as the features seen in NGC4921 \citep{Kenney+15}. Another possibility is that these are re-accreted stellar clumps, originally formed in the inner tail, but that did not reach the escape velocity and fell back onto the galaxy, forming stellar streams that occupy a thick disk or halo \citep{Abramson+11}. Such re-accretion of stars formed in ram pressure stripped tails has been found in simulations \citep{Tonnesen+12}.

\subsection{Dust Extinction in Main Body \& Tail}

\begin{figure*}
	\plottwo{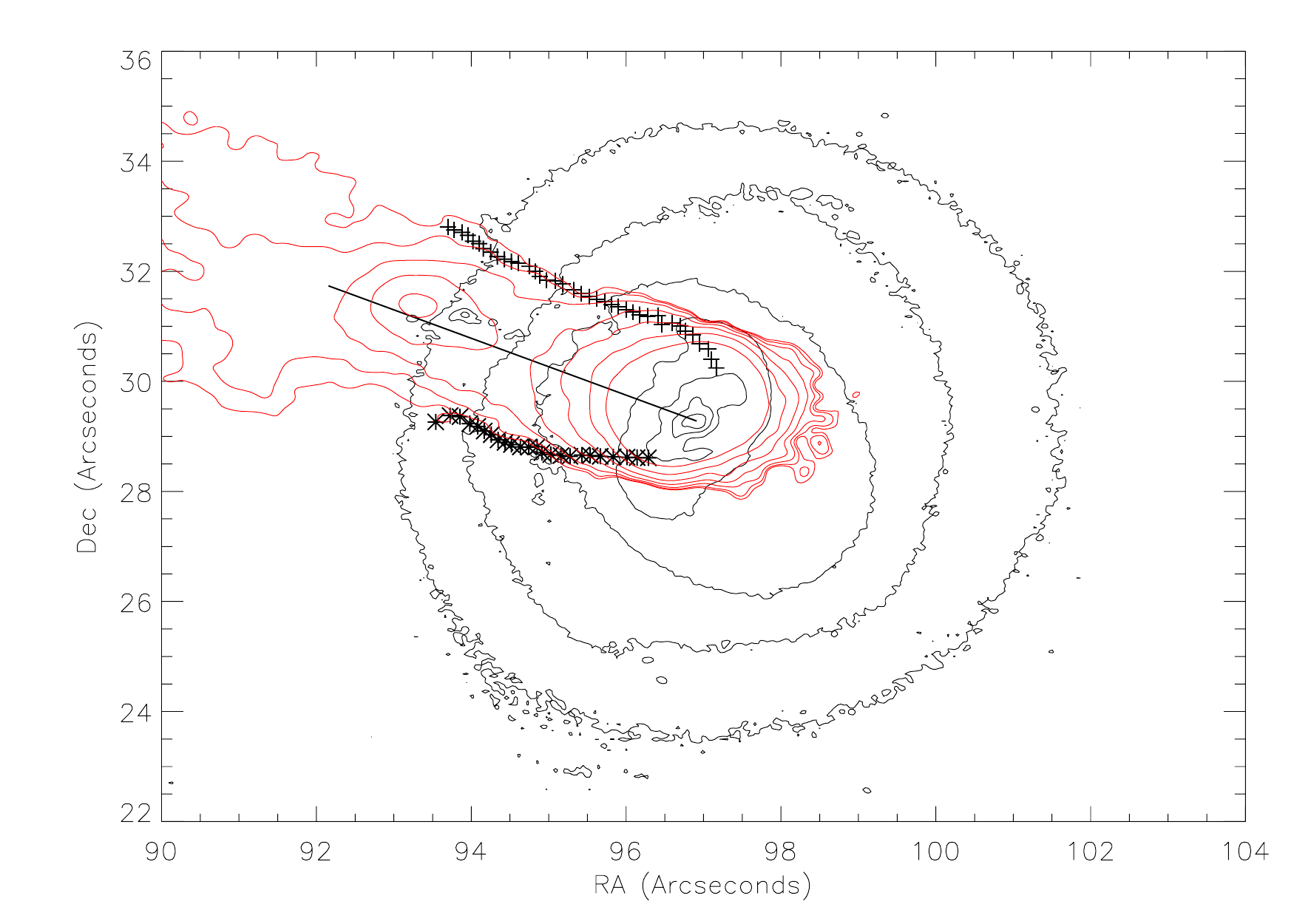}{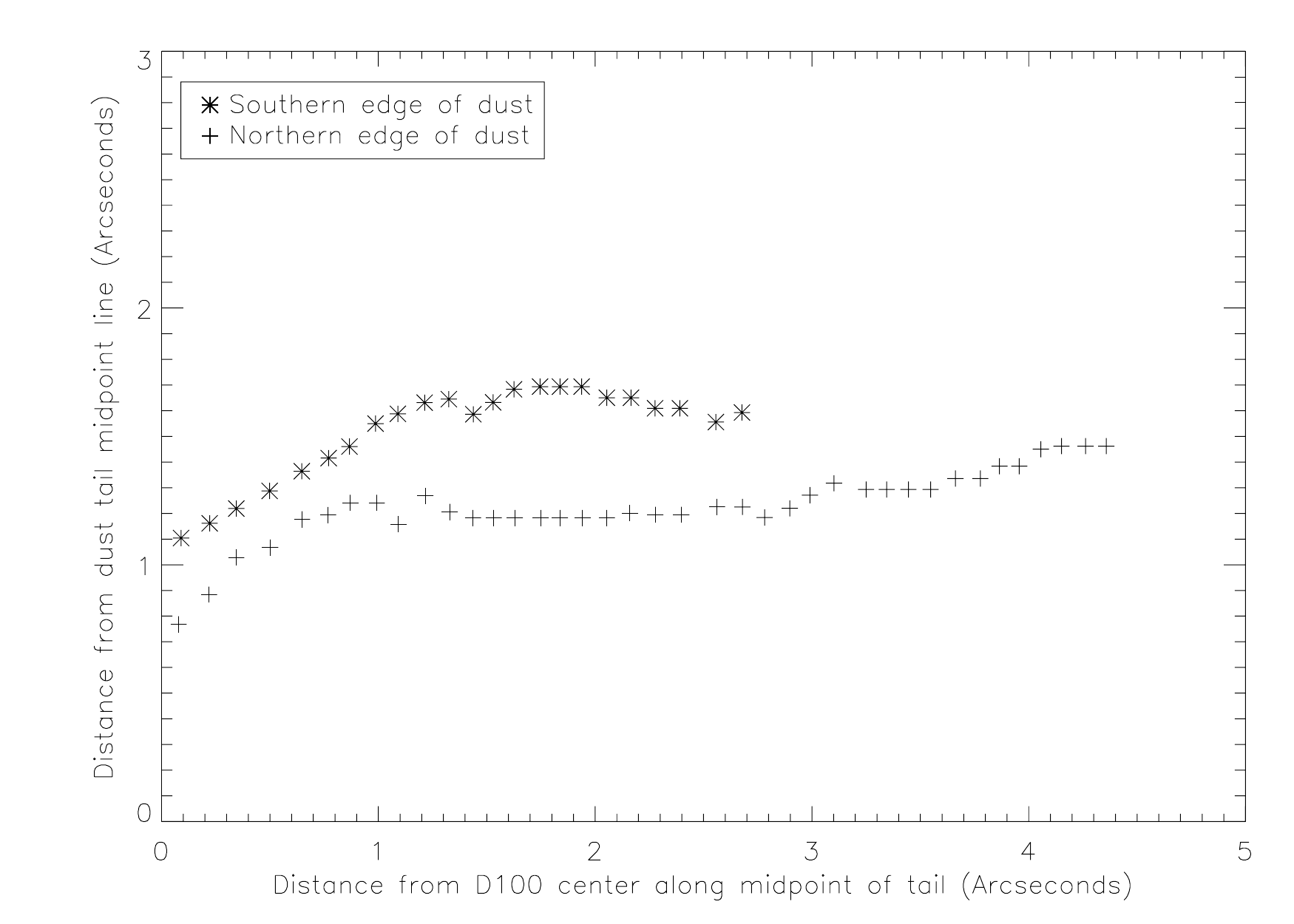}
	\caption{On the left, F814W contours of D100 are shown in black; the contour levels vary from 18.9 to 20.7 mag/arcsec$^2$ by increments of 0.36 mag/arcsec$^2$. H$\alpha$ contours are shown in red; the contour levels vary from 12.5 to 800 $\times \, 10^{-18}$ erg s$^{-1}$ cm$^{-2}$ arcsec$^{-2}$ by a factor of two for each contour level. It should be noted that the H$\alpha$ contours are unreliable in the central 1'' due to imperfect continuum subtraction. Outside of the central 1'' there is good agreement between the edge of the visible dust and the H$\alpha$.  The line is drawn from the center of the galaxy in the direction of H$\alpha$ tail, parallel to the N filament of dust. On the right, distance from the tail line drawn on the left, to the points at the edges of the dust tail for both the top and bottom of the tail. The asterisks mark the southern visible edge of the tail in extinction, and the plus signs the northern visible edge. Past the furthermost points from the nucleus, the surface brightness of the galaxy is too low to visibly detect any dust extinction from the surroundings. It can be seen the tail appears to broaden rapidly near the circumnuclear region, but maintain a near uniform width further out.}
	\label{fig:dustpoints}
\end{figure*}

The region of strongest extinction is the circumnuclear region of the galaxy, shown by \citet{Caldwell+99} to be a starburst region, based on optical spectroscopy. The strong extinction seems to have a well-defined extent, measuring $\sim$ 2'', or $\sim$ 0.93 kpc in diameter through the nucleus, along the major axis. The visible dust tail extends outward toward the NE at PA = 178$^\circ$ with respect to the major axis, from the center of D100 to a distance of at least 6.5'' (Figure \ref{fig:dustpoints}). It is clear that this dust lies above the disk plane, and thus is an extraplanar feature. The spiral structure shows that the galaxy rotates clockwise, and the N side of the galaxy has been shown to be approaching \citep{Jachym+17}, thus the eastern side of the disk is the far side. Dust in the galaxy near the disk plane on the far side of the disk does not create strong extinction, so strong dust extinction viewed toward the far side of the disk must originate from extraplanar dust closer to us than the disk midplane.

The extinction in the tail has a well defined outer envelope, with remarkably smooth and straight edges. In Figure \ref{fig:dustpoints}, we plot the width of the tail, measured by dust extinction, versus distance along the minor axis. The width of the tail, measured by dust extinction, ranges from 2'' to 3'' at a distance of 2.5'' along the minor axis. The width increases by $\sim$ 50\% within 1'' along the minor axis, then widens less, increasing only 20\% out as far as 3'' on the southern edge, and 4'' on the northern edge. We note, however, that our measure of the initial tail broadening around the center of the galaxy may be biased by the fact that dust remaining in the disk will have weaker extinction. Thus, while the dust extinction in the tail is seen much more strongly, as it is in front of the disk, dust in the disk of the galaxy near the circumnuclear region may still be present.

Outside the circumnuclear region (the inner $r \sim$ 1'' of the galaxy) the dust tracks remarkably the outskirts of the H$\alpha$ tail, as seen in Figure \ref{fig:dustpoints}. However, since the resolution of the H$\alpha$ image is much poorer than that of the HST images (0.75'' vs 0.11''), there may be more dust at the edges of the tail, while H$\alpha$ emission concentrated more near the center of the tail is artificially smoothed outward to the edges. Interpreting the width of the H$\alpha$ tail in the circumnuclear region of the galaxy is also difficult, as separating the H$\alpha$ emission from continued star formation in the disk from the emission associated with gas excited in the tail is not possible. Furthermore, the H$\alpha$ image has imperfect continuum subtraction because of the dependence on the continuum color \citep{Yagi+10, Spector+12}. The effect is large where the continuum is bright\footnote{For example, there is some excess emission in the narrowband image around the center of D99. However, a spectrum from the eBOSS survey \citep{Dawson+14} as part of SDSS DR14 \citep{Abolfathi+17} shows that the H$\alpha$ line is not detected in this galaxy. This indicates the apparent emission seen in the narrowband image is not H$\alpha$, but residual broadband continuum.}.

There is more extinction to the N of the minor axis than to the S, so the morphology is clearer to the N. The N side of the dust tail has a well-defined, fairly smooth and straight outer envelope, which extends relatively unbroken for at least 4.4''= 2.1 kpc. A band of strong extinction at the edge has a uniform width of 0.5''= 237 pc. In places it appears as 2 parallel lanes, i.e., doubled, with variable extinction in between and along the lanes. There is less extinction in the S edge of dust tail so its morphology is not as clear, but it seems to have  a well-defined and straight outer edge. In between the N tail edge and the minor axis is a more irregular distribution of strong extinction.

The smooth and straight tail edges, the linear substructure of some dust features, and the small amount of tail broadening, are remarkable and unexpected from simple stripping models. We discuss the implications of this further in Section 4.

\subsection{Evidence for Ram Pressure Stripping in D100 and its neighbors D99 and GMP2913}

Here we discuss constraints on the star formation histories of D100 and its two nearest apparent neighbors, D99 and GMP2913, and evidence on ram pressure stripping and tidal interactions for the galaxies. There is evidence from the HST data that all three galaxies have experienced ram pressure stripping. Outside-in quenching as a result of ram pressure stripping in clusters is a well known prediction from basic physics, and from simulations \citep{Boselli+06, Kapferer+09, Tonnesen+10},  but high resolution studies of the timescale of the radial progression of quenching are a recent and active area of study (e.g. \citet{Pappalardo+10, Abramson+11, Merluzzi+16, Fossati+18}).

\subsubsection{D100}

\begin{figure*}
	\plotone{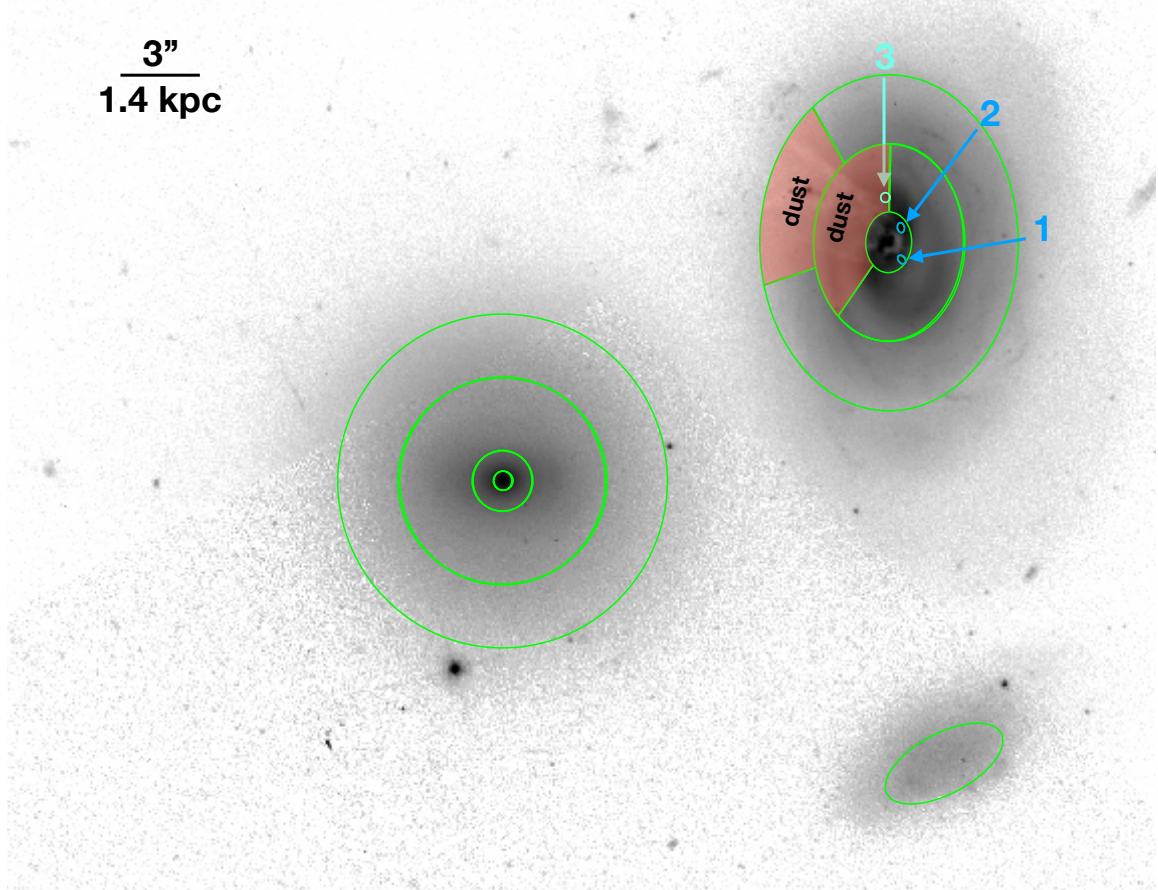}
	\caption{Regions for color analysis; the background is the F475W image. For the regions in D100, the sections of the galaxy with obvious dust extinction near the tail, and around the circumnuclear region, are excluded from the elliptical apertures. Inner regions selected for their especially strong F275W emission are labeled with numbers that correspond to their labeling in the model plot in Figure \ref{fig:burstmodel}.}
	\label{fig:starburst_regions}
\end{figure*}

Our high resolution HST data allows us to quantitatively measure the radial gradient of star formation quenching. We select various iso-color regions of the three galaxies at varying radii. Regions selected are shown in Figure \ref{fig:starburst_regions}.

\begin{figure*}
	\plotone{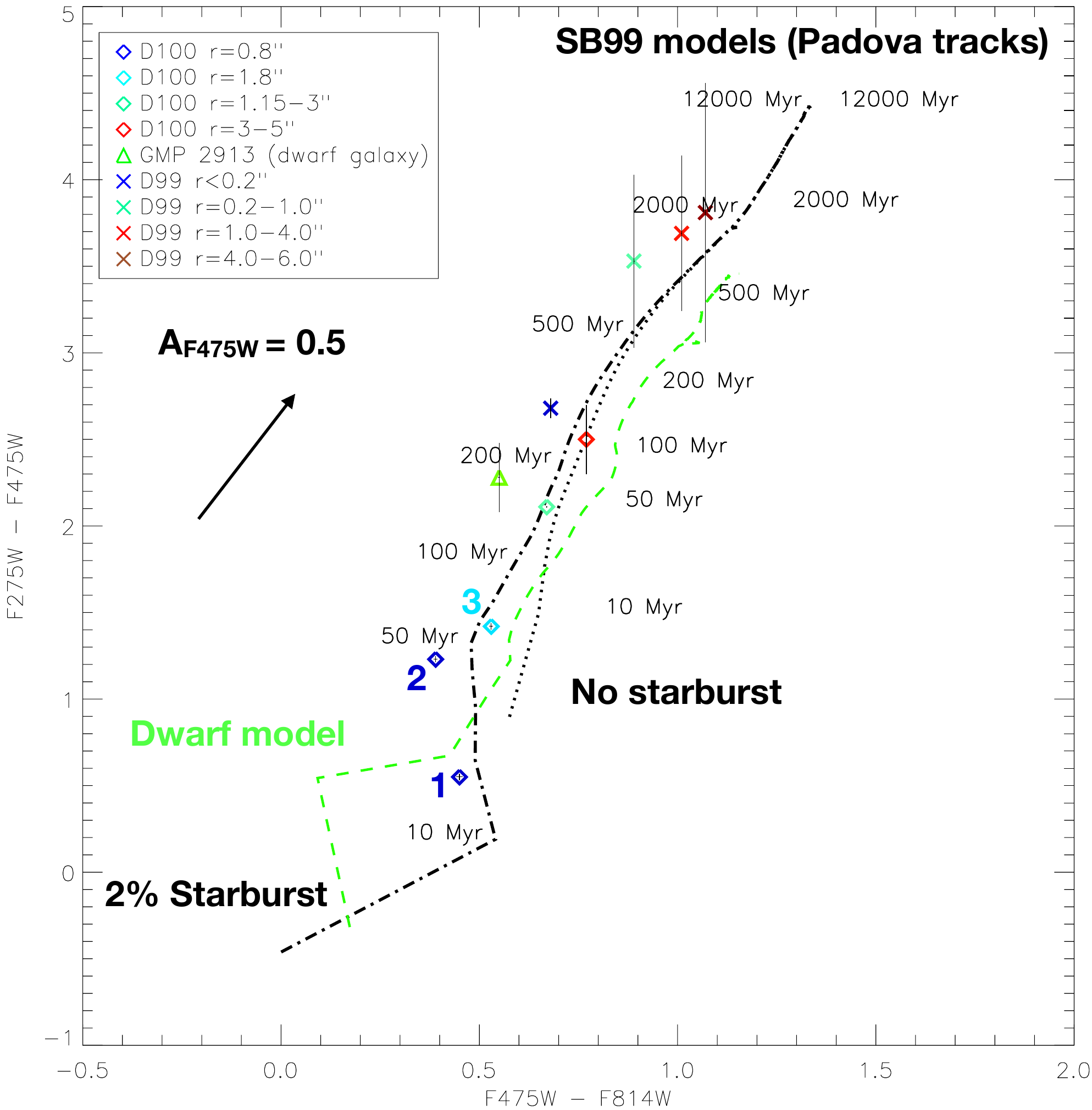}
	\caption{A color-color plot of isocolor regions of the three galaxies in this study. The lines show stellar population models generated with Starburst99. The black lines show models for solar metallicity; they also have corresponding age labels. The black, dot-dashed line shows a truncation model with a 2\% burst, while the black, dotted line shows a simple truncation model. The green line shows a 2\% burst model with a metallicity of 0.2 solar abundance, to match that expected for the dwarf galaxy, GMP 2913. There is only one annulus drawn for the dwarf galaxy because it has uniform colors throughout. While we show for reference an extinction vector (for an arbitrary $A_{F475W}=0.5$ mags), note that we don't believe any of these regions contain significant dust as the obviously dusty portion of D100 was excluded, and D99 \& GMP 2913 have likely been stripped of all their ISM.}
	\label{fig:burstmodel}
\end{figure*}

In order to analyze the observed colors of several regions of the three galaxies, we make use of the Starburst99 stellar population models \citep{Leitherer+99, Vasquez+05}. Our input parameters for this stellar population assume a standard Kroupa initial mass function (IMF) and utilize Padova isochrones \citep{Bressan+93, Fagotto+94a, Fagotto+94b, Girardi+00}, with added thermally pulsating AGB stars. Both D100 and D99 have stellar masses near $\sim$ 4 $\times \, 10^9$ M$_\odot$, and thus are expected to have metallicities of 12 + log(O/H) $\sim$ 8.7, about solar \citep{Sanchez+17}. The dwarf galaxy GMP 2913 has a much smaller mass (1.8 $\times 10^8$ M$_\odot$), and thus, based on the mass-metallicity relation \citep{Tremonti+04}, we use a 0.2 solar metallicity model for that galaxy. We use the output from Starburst99 to construct the SED of a stellar population with a truncated star formation history. This history assumes star formation started 12 Gyr ago, proceeded at a constant rate, then was abruptly truncated (Figure \ref{fig:burstmodel}, dotted line). Additionally, we construct a SED for a truncated star formation history where star formation momentarily increased at the time of truncation, such that 2\% of the stellar mass formed at the time of truncation -- an increase in SFR by a factor of $\sim$ 25 for a timestep of 10 Myr (Figure \ref{fig:burstmodel}, dash-dotted line). Using the SED outputs from these models, we extracted broadband colors for the F275W, F475W, and F814W bands, which we compare with our selected regions.

\begin{figure*}
	\plotone{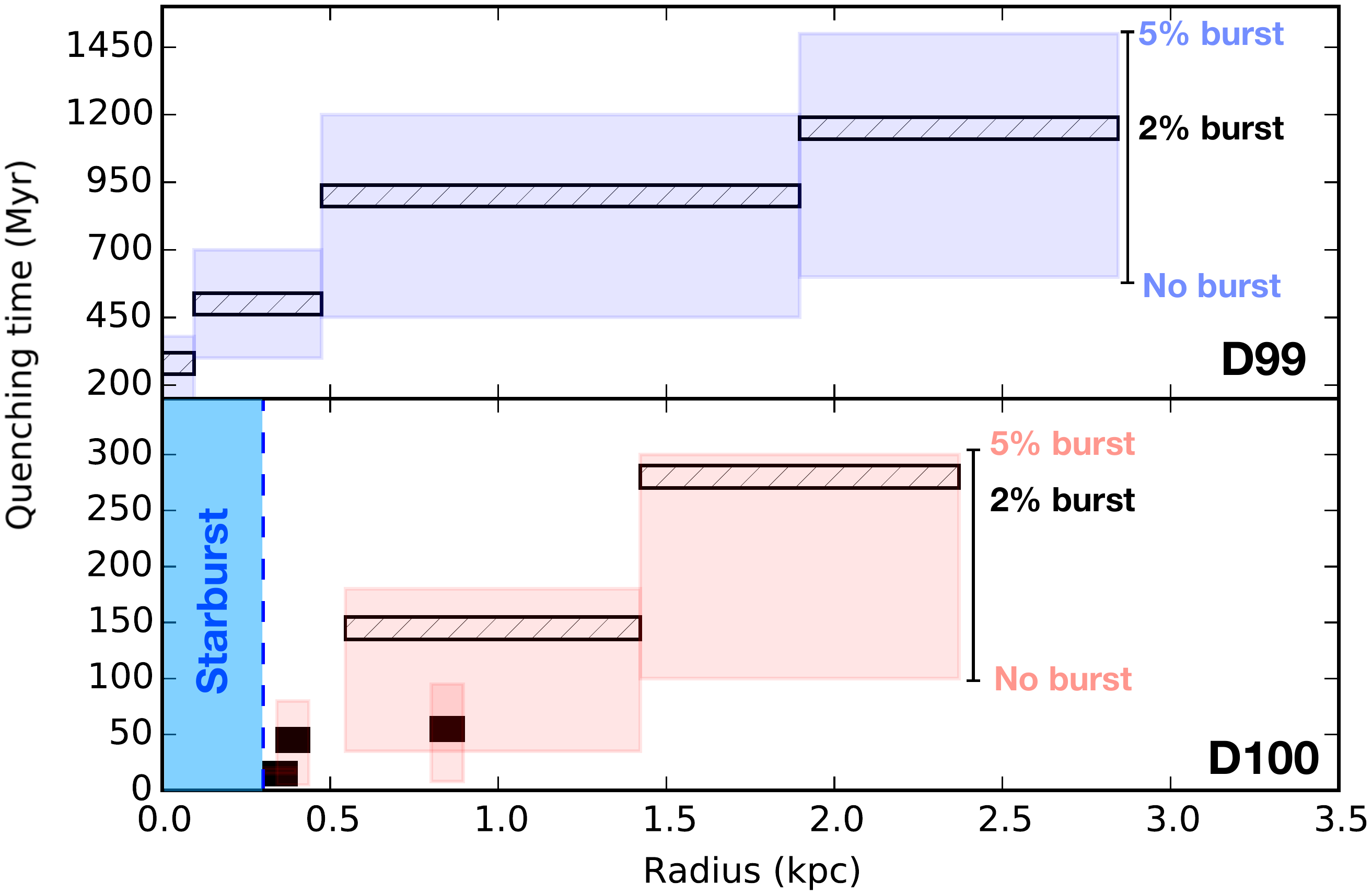}
	\caption{Quenching time versus radius for regions in D100 and D99 marked in Figure \ref{fig:burstmodel}. The cross-hatched boxes show the estimate of the quenching time based on our SB99 model with a 2\% burst. The solid black boxes are the smaller regions in D100, marked as 1, 2 \& 3 in Figure \ref{fig:starburst_regions} with the quenching time also estimated from the 2\% burst model. The shaded boxes show the upper and lower limits on the quenching time based on the 5\% burst, and the simple truncation model (the simple truncation model, does not extend to the bluest measured region, region 1, so here the lower bound is the the 0.4\% burst model). In the lower plot, the vertical dashed line shows the estimated stripping radius of D100, within which star formation is still ongoing. Both galaxies show clear evidence for outside-in quenching.}
	\label{fig:quenching_gradient}
\end{figure*}

In D100, the radial extent of the dust extinction and the H$\alpha$ emission indicates that ongoing star formation is confined to the central $r \sim 0.8$''. Beyond this radius, there seems to be no ongoing SF and no dust, and here we use the HST colors to estimate quenching times. Our data show a clear radial color gradient from within $r \sim 0.8$'' out to between $3$''$ - \, 5$'', evidence of outside-in quenching. The quenching times we derive from this color gradient are shown in Figure \ref{fig:quenching_gradient}. From our color measurements alone we cannot break the degeneracy between burst strength and quenching time. Some of the colors could be fit with either a simple truncation model, or with a truncation plus burst model with an older quenching time. However we can break this degeneracy using the spectroscopy results of \citet{Caldwell+99} for the region outside $r$ $\sim$ 3'', for which they found  a postburst population with a quenching time of $\sim$ 250 Myr. We find that a model with a burst strength of 2\% results in a quenching time of $280 \pm 20$ Myr in this region, so we adopt this burst strength as our preferred model for all regions. However we also calculate quenching times for different burst strengths (0.4\% and 5\%), to provide an estimate of the systematic uncertainty in quenching times. Within $r=0.8$'', the two regions tested are so blue they require a starburst, and cannot be explained with the simple truncation model. For $r=1.15$''$ - \, 3$'', we derive a quenching time of $\sim 145^{+35}_{-110}$ Myr based on our 2\% starburst burst model.

The star formation radial quenching profile of D100 shows an interesting correspondence with the tail width and age, which is discussed further in Section 4.

\subsubsection{D99}

\begin{figure*}
	\plotone{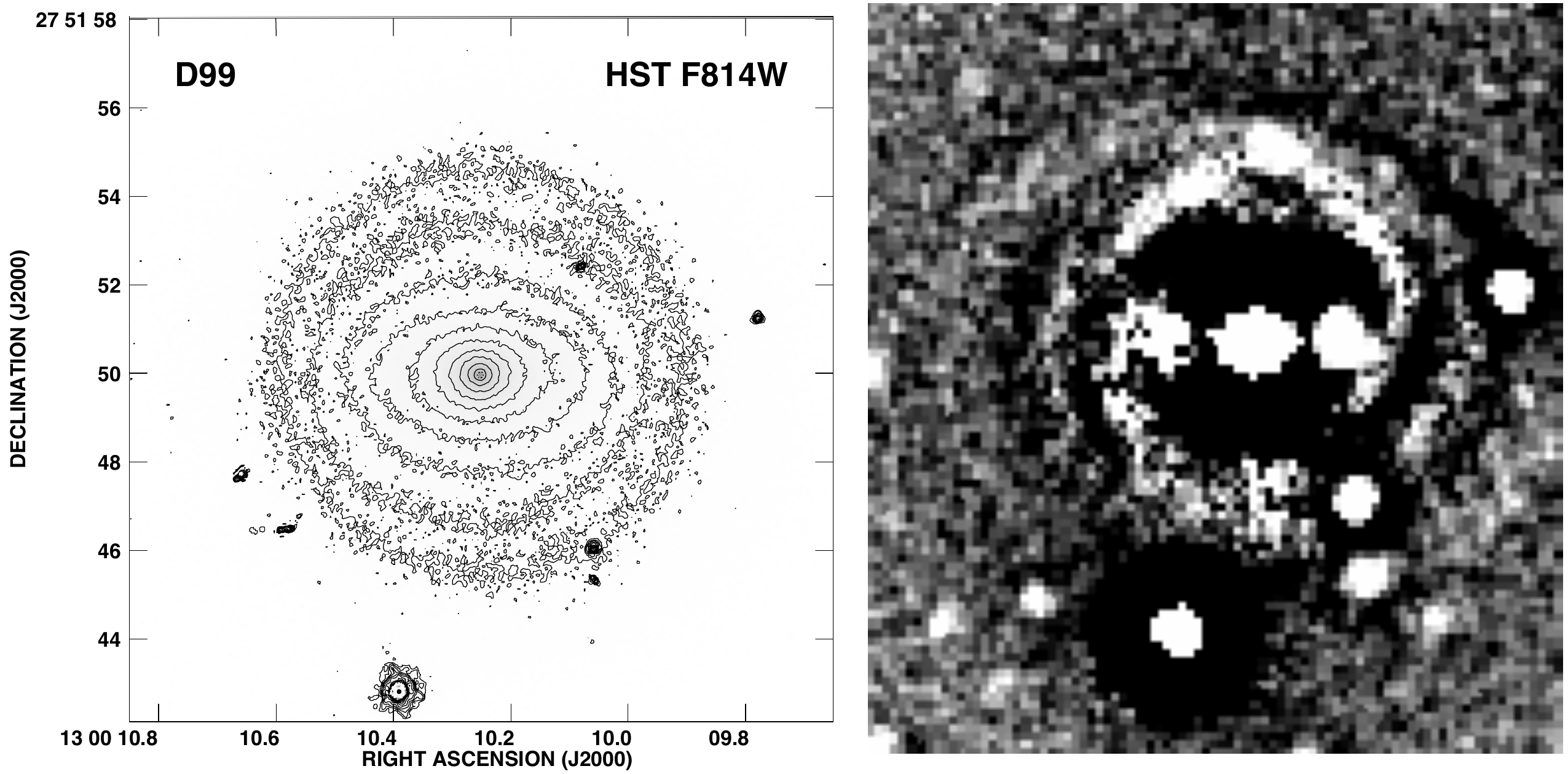}
	\caption{On the left, an isophotal contour plot of D99 from our HST F814W data. The contour levels vary from 21.4 to 16.3 mag/arcsec$^2$ by increments of 0.36 mag/arcsec$^2$. On the right, an unsharp-mask image to highlight sub-structure, generated from the Subaru $R$-band image of D99. Both figures show a faint bar or lens, with the unsharp masked image revealing two spiral arms. This suggests that D99 is a fully stripped spiral.}
	\label{fig:D99_structure}
\end{figure*}

The nearby (on the sky) galaxy D99 (also called GMP 2897) is located 0.29 arcminutes SE of D100 (Figure \ref{fig:sidebyside}). D99 is nearly face-on, and similar in radial size to D100, with an absolute magnitude of $M_{F814W}=-18.97$, compared to the absolute magnitude of D100, $M_{F814W}=-19.65$. Its stellar mass is also quite similar, 4.7 $\times 10^9$ M$_\odot$ from the WiscM11 SED model \citep{WiscM11}. As noted previously, it is very unlikely to be physically associated with D100, as it differs in radial velocity by $\sim$ 4500 km s$^{-1}$. The velocity of D99 ($\sim$ 9900 km s$^{-1}$) at a projected distance of 240 kpc is on the extreme end of bound Coma galaxies, but theoretical predictions for bound cluster members presented in \citet{Kent+82}, and a more recent empirical sample in \citet{Kadowaki+17} both support D99 being a bound cluster member. We present an isophotal analysis and unsharp mask image in Figure \ref{fig:D99_structure}, which highlights some similarities with D100. We hypothesize that D99 is a similar galaxy to D100 at a later evolutionary stage of the same ram pressure stripping process.

HST light profiles and the unsharp mask image reveal a bar/lens in the center, and two faint spiral arms, as well as a bulge + disk radial light profile. These morphological characteristics, along with the lack of H$\alpha$ emission from both the Subaru imaging data and SDSS spectroscopy, strongly suggest that D99 is a stripped spiral galaxy, not yet converted fully to a ``red and dead" lenticular galaxy via RPS. Our stellar population models in Figure \ref{fig:burstmodel} can't distinguish between a simple truncation and a truncation with a burst. However, spectroscopy by \citet{Caldwell+99} suggests a $\sim$ 1 Gyr old poststarburst population beyond $r=1$''; this age agrees with both our 2\% and 5\% burst model. We then find that the color gradient in D99 corresponds to a quenching time (with the 2\% model, with the lower bound being the truncated model, and the upper bound the 5\% burst model) of $\sim 280^{+100}_{-135}$ Myr at $r$ $<$ 0.2'', $\sim 500^{+200}_{-200}$ Myr at $r = 0.2$''$ - \, 1.0$'', $\sim 900^{+300}_{-450}$ Myr at $r=1.0$''$ - \, 4.0$'', and $\sim 1150^{+350}_{-550}$ Myr at $r = 4.0$''$ - \, 6.0$''. This data is plotted in Figure \ref{fig:quenching_gradient}, and supports a trend of outside-in quenching consistent with RPS. Furthermore, the large difference in quenching time between the central region of D99 and the outskirts suggests that it may take several hundred Myr more to quench the central regions of D100 completely, if stripping proceeds at a similar rate.

Both the morphology, and the outside-in radial quenching gradient strongly suggest that D99 is very much like D100, but caught at a later evolutionary stage, $\sim 280^{+100}_{-135}$ Myr after the last nuclear gas was stripped.

\subsubsection{GMP 2913}

GMP 2913 is a dwarf irregular galaxy, with an absolute magnitude of $M_{F814W}=-16.65$, about 3 magnitudes fainter than D99 and D100. Assuming a stellar M/L ratio of 1, the stellar mass of this galaxy is 1.8 $\times 10^8$ M$_\odot$, only $\sim$ 4\% the mass of D100. This corresponds to an expected metallicity around 0.2 solar \citep{Sanchez+17}. It is located 0.33' = 9.37 kpc from D100 on the sky, and has a velocity within $\sim$ 200 km s$^{-1}$ of D100 \citep{Yagi+07}, thus it is possibly gravitationally bound to D100, although the velocity in the plane of the sky is unknown. There is evidence that GMP 2913 is, or has recently in the past, gravitationally interacted with D100. Its irregular structure, and the irregular southern side of the disk of D100 suggest a tidal interaction. It has quite uniform colors, with F475W-F814W $\sim$ 0.5 throughout, but very low level, diffuse F275W, and no visible dust extinction. In the spectrum obtained by FOCAS/Subaru \citep{Yagi+07}, there is no H$\alpha$ emission, and is clear H$\beta$ and H$\alpha$ absorption. Thus, there is no ongoing star formation; however, since no shorter wavelength information is available to search for strong H$\delta$ absorption, it is unclear if GMP 2913 is also a post-starburst galaxy. With a 2\% burst model we estimate its quenching time is $\sim 300^{+100}_{-150}$ Myr, with a significant uncertainty as it does not fall very close to the model colors in Figure \ref{fig:burstmodel}.

If GMP 2913 and D100 are indeed bound to each other, they are likely orbiting together through the cluster, and have experienced RPS at about the same time. It is likely that all the gas was stripped out of GMP 2913 first and more rapidly, as it is a lower mass galaxy than D100, with a weaker gravitational potential. Rapid stripping of the entire galaxy would explain the uniform color of the dwarf galaxy.

\subsection{Stellar sources in the tail}

\subsubsection{Estimating ages and masses of stellar clumps}

The ram pressure stripped tail of D100 has been shown to contain large amounts of molecular gas \citep{Jachym+17}, required for the formation of stars, accompanying the prominent H$\alpha$ emission. However, it is not known how much of this H$\alpha$ emission is due to young stars.

\begin{figure*}
	\plotone{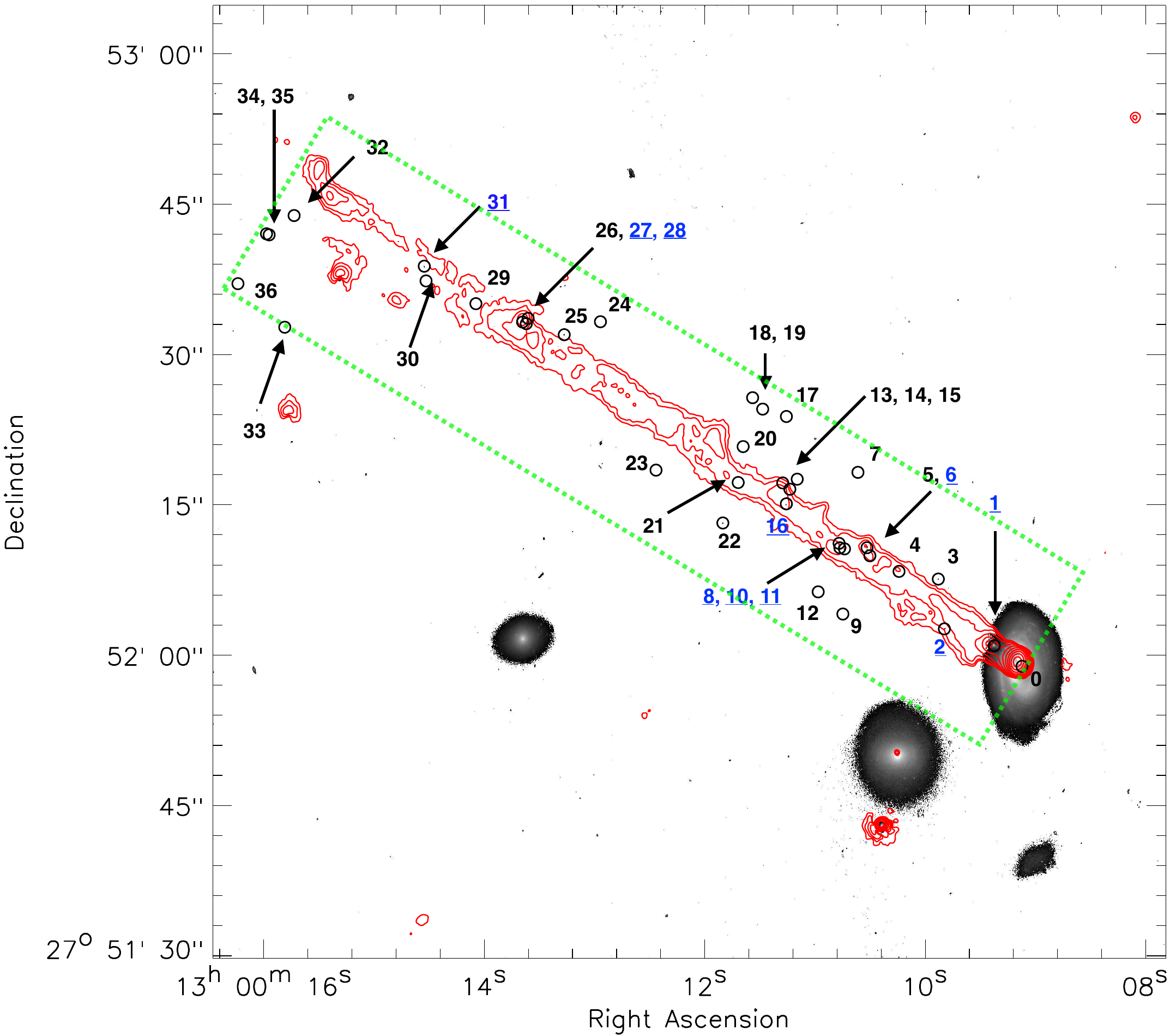}
	\caption{Sources in the tail region of D100 over-plotted on a greyscale F475W image. In red, contours from a smoothed H$\alpha$ image, the contour levels vary from 4.8 to 160,000 $\times \, 10^{-18}$ erg s$^{-1}$ cm$^{-2}$ arcsec$^{-2}$ by a factor of two increment between contour levels. The green rectangle shows the search area. Sources are numbered, with those shown in blue having positive F275W detections.}
	\label{fig:rectangle}
\end{figure*}

In order to quantify the amount of star formation in the tail of D100, we made a catalogue of sources in the tail region. We searched for sources within a rectangular region, with the length set as the extent of the H$\alpha$ tail, $\sim$ 105'', and the width set as the length of the apparent optical disk of D100 ($\sim$ 20''). We set the search area as the width of the optical disk of the galaxy such that we could find evidence of both stars in the current gas tail, and any stars that may have formed when the outskirts of the galaxy were stripped. We objectively identified surface brightness peaks that could be stellar sources in the tail of D100 using Source Extractor \citep{Sextractor+96} operating on the F475W band, as it had the highest S/N. Our detection criteria were set to find sources with four contiguous pixels above 3$\sigma$ significance. Then, utilizing the dual-imaging mode of Source Extractor, the apertures from the F475W band were applied to the F814W and F275W band. The list of sources was refined by removing obvious background galaxies, identified by their morphology (such as clear spiral arms, or exponential disks). However, interlopers that were more poorly resolved may still remain intermixed with stellar sources in the tail. The locations of the resulting detections are shown in Figure \ref{fig:rectangle}, and includes a total of 37 identified sources, with about half found in the H$\alpha$ tail, and the other half outside. All sources are detected in the F814W, however, only 10 sources are detected in the F275W filter, and all of these ten sources were found within the extent of the H$\alpha$ tail. Sources which did not have a F275W detection were assigned an upper limit value. This value was based on the dimmest visible object in the F275W image, with $m_{F275W} = 28.3$, about 2$\sigma$ above the background.

\begin{figure*}
	\plotone{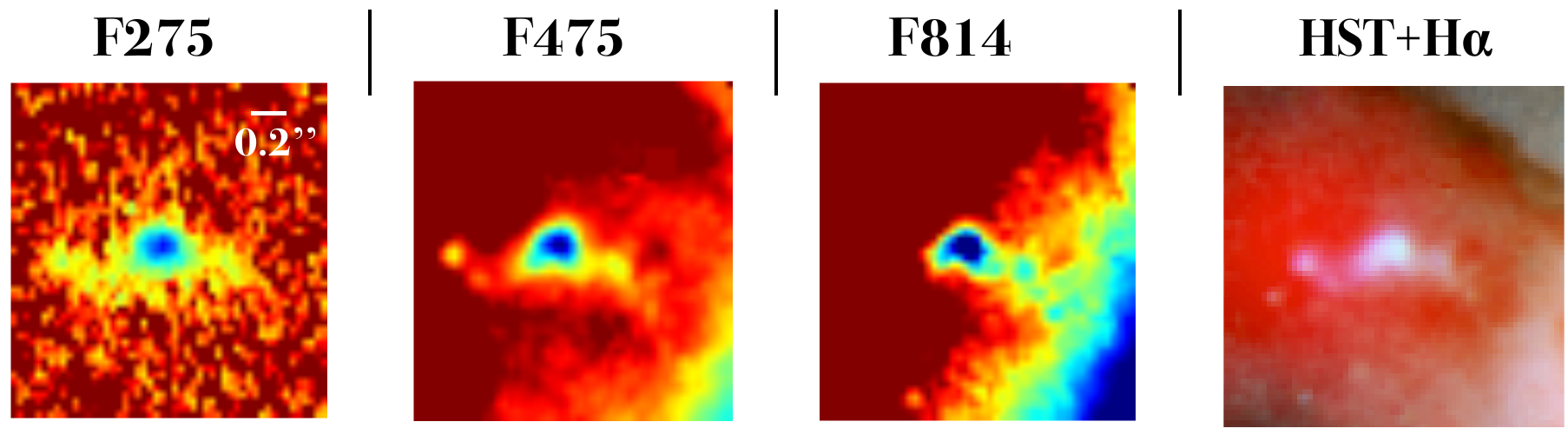}
	\caption{Images of source 1, located at the base of the tail. From left to right: HST F275W, F475W, F814W, and F475W+F814W with H$\alpha$ in red from ground-based Subaru observations \citep{Yagi+10} overlayed. Distinct features of the star clump can be seen, such as two quasi-arms of stars, with a concentration seen in the F275W at the easternmost tip, the brightest point outside the center of the region. It should be noted that in the F814W image, there is significant background emission from the disk of the galaxy, while in the F275W image, there is no measurable background, as stars in the disk here are much older than the star forming complex.}
	\label{fig:numberedsources}
\end{figure*}

\begin{figure*}
	\plotone{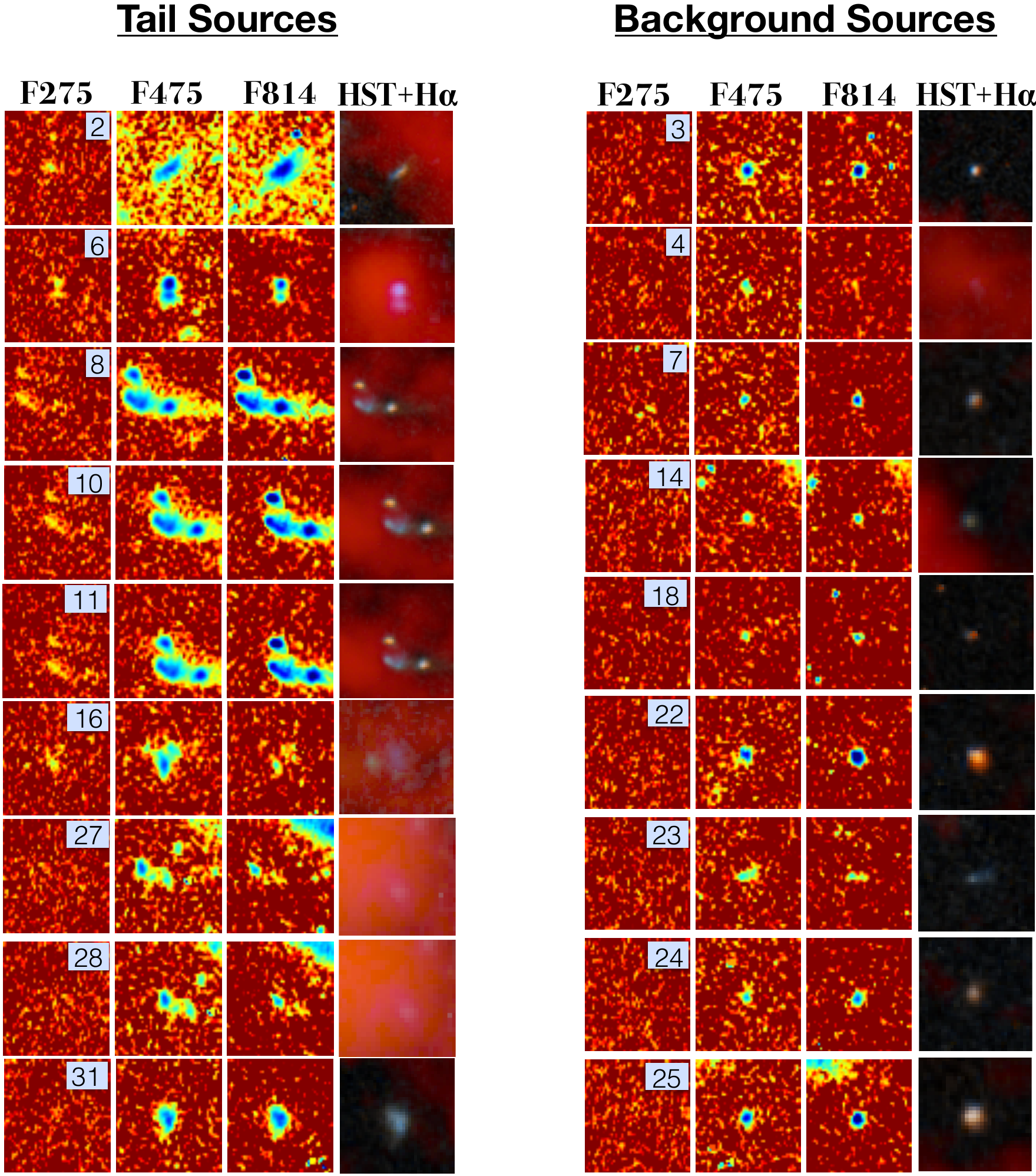}
	\caption{On the left, 2 arcsecond cutouts of all F275W detected sources in, from left to right, F275W, F475W, F814W, and F475W+F814W with H$\alpha$. On the right are some examples of sources with no F275W detections that are likely background sources. All of these sources are also shown labeled in Figure \ref{fig:rectangle}. The stamps are centered on the source that is numbered.}
	\label{fig:stamps}
\end{figure*}

Images of all F275W detected sources, and a selection of sources of varying morphologies with no F275W emission are shown in Figures \ref{fig:numberedsources} \& \ref{fig:stamps}. Figure \ref{fig:numberedsources} shows the brightest source, source 1, the only one detected with enough sensitivity to show clear substructure in all bands.

\begin{figure*}
	\plotone{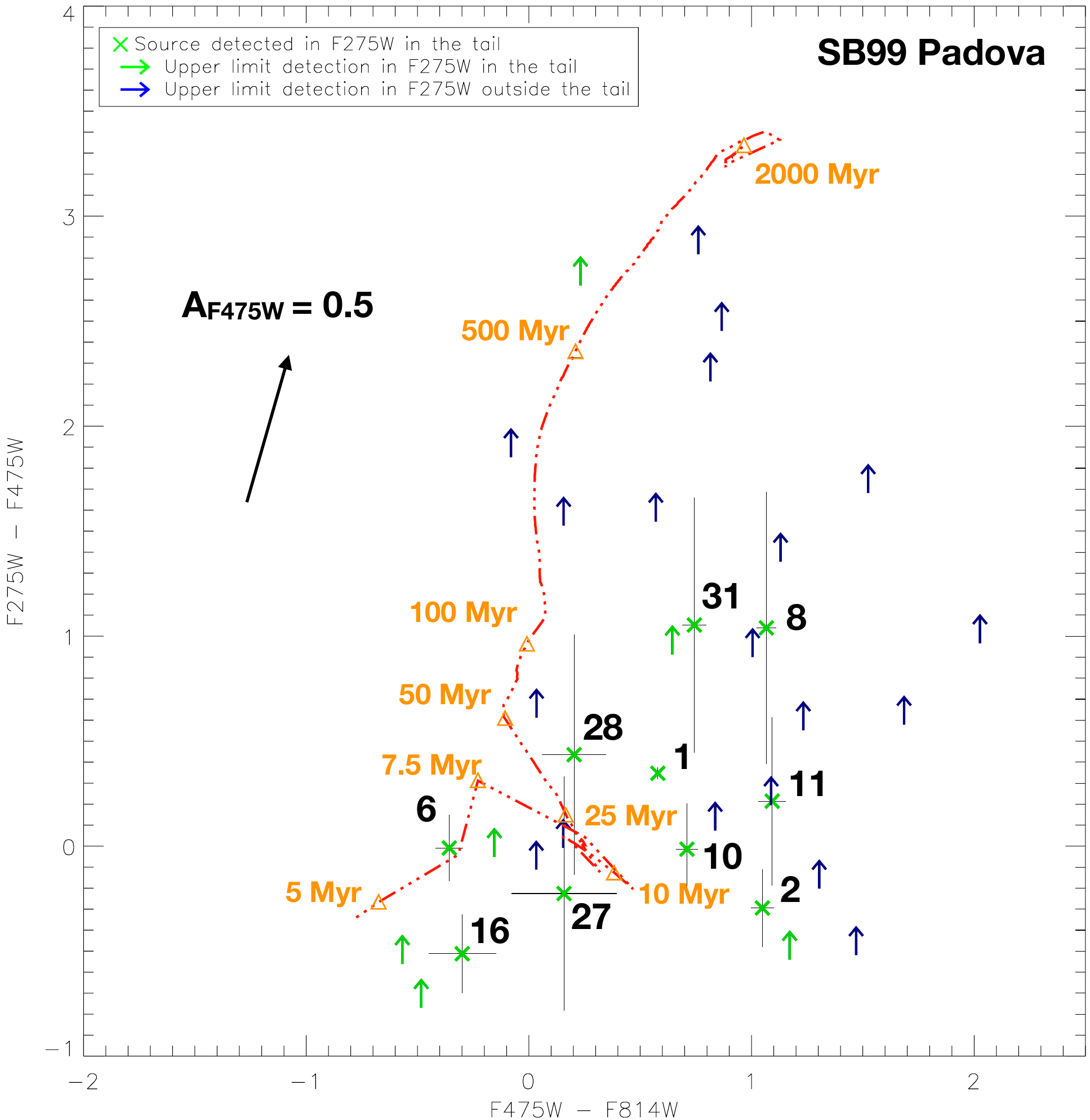}
	\caption{Color-color diagram of sources detected in the F275W filter, all of which are found in the tail, and limit sources both inside and outside the tail. Also shown is our SSP model track generated with SB99 with Padova tracks. The model is generated for a population with solar metallicity (Z=0.02), the labeled model ages are shown as open triangles. Sources are labeled with their identifier from Figures \ref{fig:numberedsources} \& \ref{fig:stamps}. The extinction vector (shown here for an $A_{F475W}=0.5$ mag, which results in a reddening value of $E(B-V) \sim 0.5$) is derived from the extinction curve generated by \citet{Dong+14}.}
	\label{fig:tracks_labeled}
\end{figure*}

\begin{figure*}
	\plotone{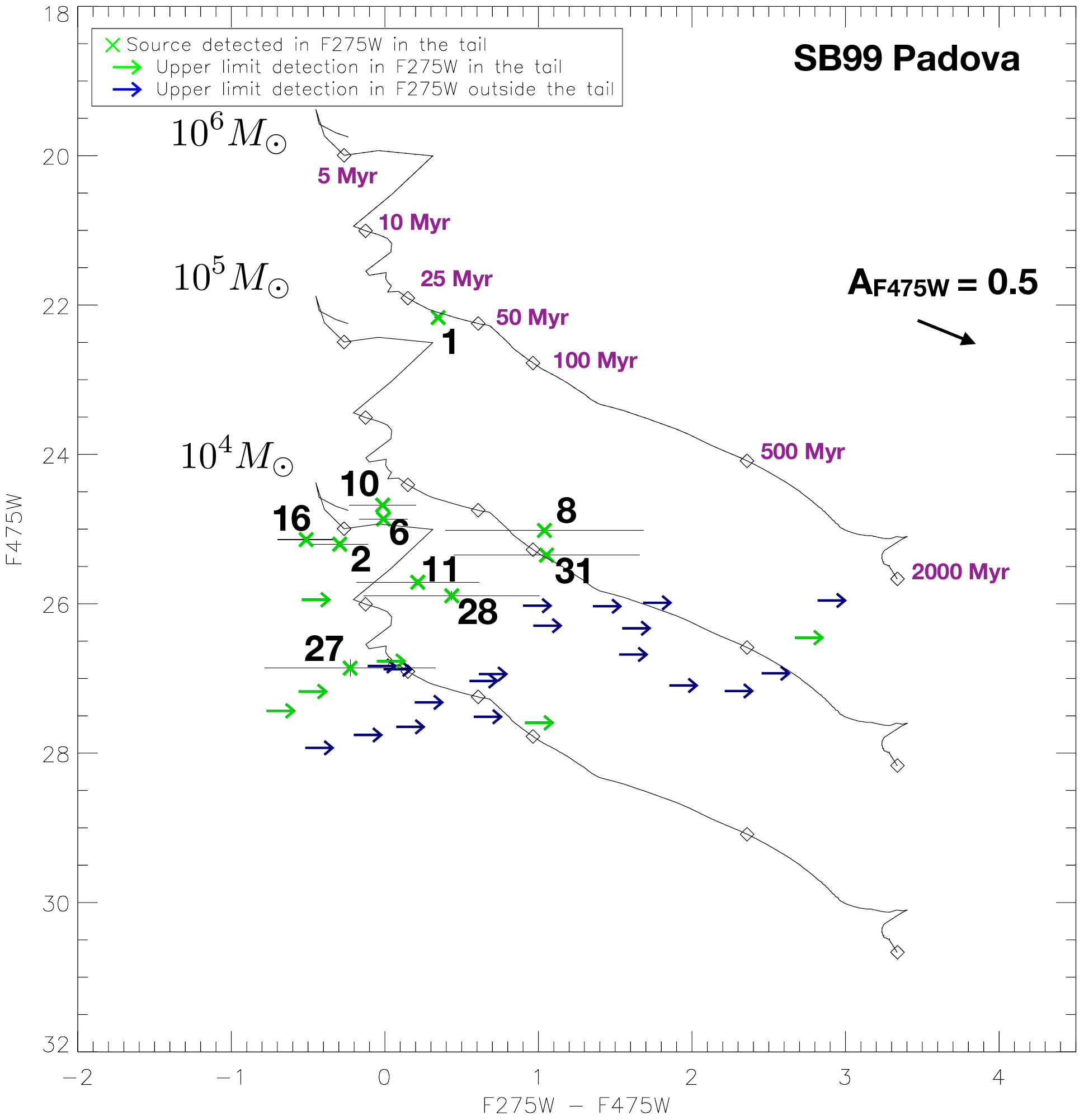}
	\caption{Color magnitude diagram of the sources in the tail region of D100 shown in Figures \ref{fig:numberedsources} \& \ref{fig:stamps}. The crosses are sources that have positive F275W detections, while the upper limit arrows correspond to sources with no detection in F275W. Symbols in green are inside an area of positive H$\alpha$ flux corresponding to the tail of D100, seen in Figure \ref{fig:rectangle}, while those in blue are outside the tail. The source that is brightest in F475W is source 1, the most visually obvious source of F275W flux shown in Figure \ref{fig:numberedsources}. The lines correspond to SSP's of a fixed mass evolving in time, generated with SB99, while labels are the age of the population at the marked open diamonds. The extinction vector is drawn for $A_{F475W} = 0.5$ mag.}
	\label{fig:colormag1}
\end{figure*}

We plot the sources in a color-color (F275W-F475W vs F475W-F814W) and color-magnitude diagram (F475W vs F275W-F475W) in Figures \ref{fig:tracks_labeled} \& \ref{fig:colormag1}. All sources with a positive F275W detection are found inside the H$\alpha$ tail, and none are found outside. Given that the ratio of the area of the tail to the area of the rectangle is 0.23, the likelihood that all F275W sources would randomly fall in the tail would be $4\times10^{-7}$. Thus, we are confident that the F275W detections correspond to young stellar complexes associated with the tail of D100, and are not extraneous background sources. If they were background sources, they should be randomly distributed within the search area, and not preferentially located in the tail.

While we earlier used Starburst99 to model an extended star formation history with an abrupt truncation, in Figure \ref{fig:tracks_labeled} we simply use the single stellar population (SSP) results from Starburst99 to estimate the ages and masses of each stellar clump, based on their color and magnitude. Given their small sizes and masses, and their relative isolation, it is likely that the stars in each clump formed in a single star forming event. Our models include the expected effects of emission from nebular lines. Strong nebular lines are only expected to fall in the F475W band, and have significant effects for sources with H II regions, i.e. younger than 10 Myr. We used the output H$\beta$ flux from the Starburst99 models, and also included the expected contribution from H$\delta$, H$\gamma$, and [OIII], using predicted line ratios from the SVO Filter Profile Service\footnote{The SVO Filter Profile Service. Rodrigo, C., Solano, E., Bayo, A. \hfill \break http://ivoa.net/documents/Notes/SVOFPS/index.html}$^,$\footnote{The Filter Profile Service Access Protocol. Rodrigo, C., Solano, E. \hfill \break http://ivoa.net/documents/Notes/SVOFPSDAL/index.html}.

The sources detected in all bands are shown in the color-color diagram Figure \ref{fig:tracks_labeled} with numerical labels. The tracks fit the observations fairly well, although there are a few sources redder in both colors than theoretically predicted. We expect there to be dust extinction in the tail, due to the prevalence of molecular gas throughout the tail, and the dust seen permeating the tail near the galaxy. For the furthest source from the tracks, source 11, to fall on the track following the slope of the extinction vector shown on the plot, 0.6 magnitude of extinction in the F475W filter would be required, according to extinction curves for HST filters from \citet{Dong+14}. In \citet{Poggianti+19}, the authors found the average extinction from the Balmer decrement in a sample of 16 ram pressure stripped tails was $\sim$ $A_V = 0.5$. Thus, our models are within a plausible amount of extinction, and we assume the offsets from the model tracks are due to solely to the error bars, and extinction, and we derive age estimates accordingly. 

For sources not detected in the F275W filter and not coincident with the H$\alpha$ tail (also shown in Figure \ref{fig:tracks_labeled}) we expect the reason they do not fall on the tracks is that they are not well modeled by SSPs. These are background sources, likely galaxies behind Coma, and thus have complex star formation histories and/or significant redshifts. Some sources undetected in F275W and coincident with the gas tail may indeed be tail stellar sources, as they fall near the model tracks. Limit sources in the tail tend to be dimmer in F475W than F275W detected tail sources, as indicated in Figure \ref{fig:colormag1}, and thus may be either too old, or too low mass to be detected in the F275W filter.

\begin{figure*}
	\plotone{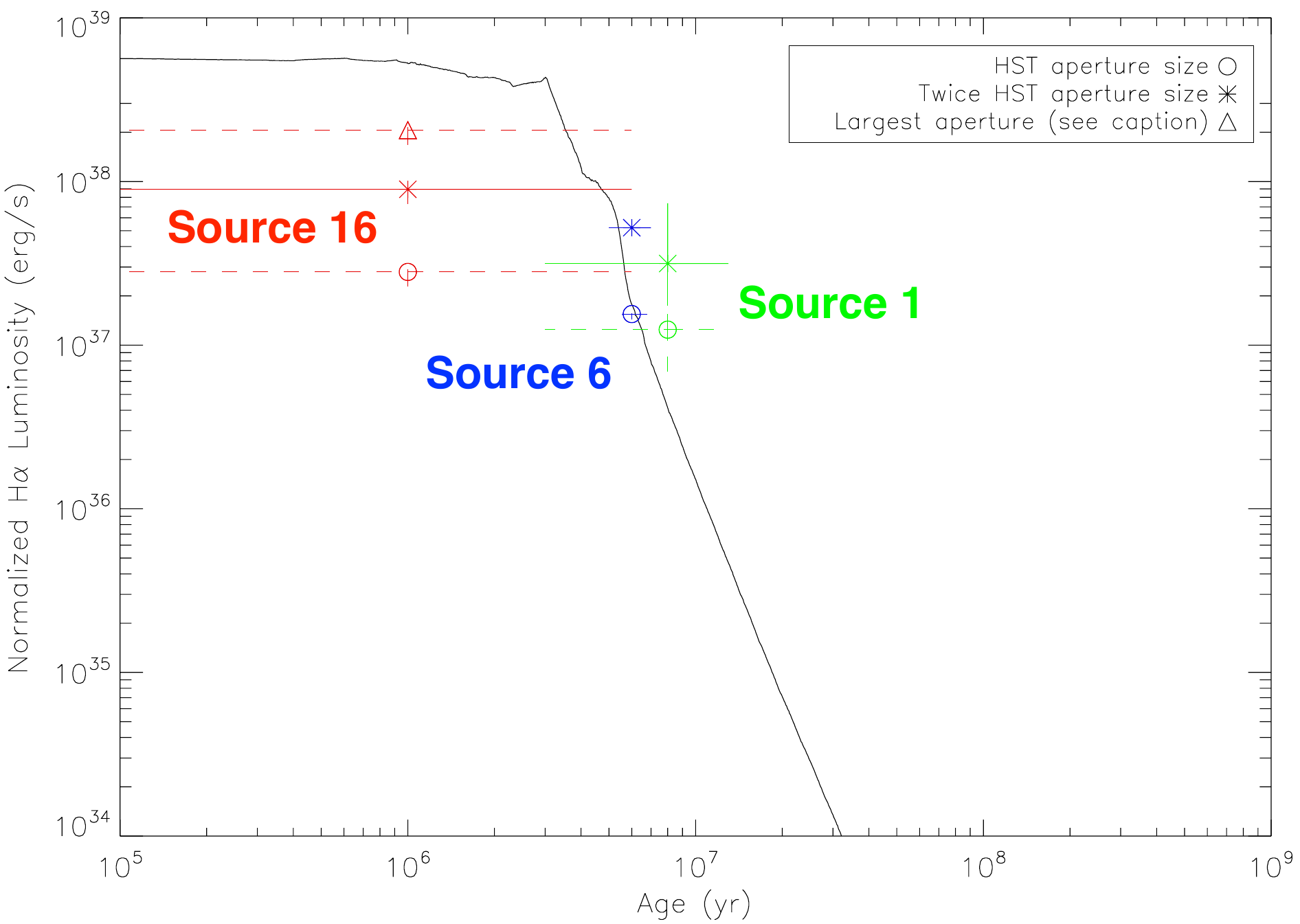}
	\caption{Comparison of the measured H$\alpha$ luminosity of three tail sources from ground based Subaru observations, and predicted nebular H$\alpha$ emission, from Starburst99 models. The H$\alpha$ luminosity of each source has been normalized to that produced by a mass of stars of 10$^4$ M$_{\odot}$, to match the model. Different aperture sizes are shown for each source, including the same size as that used to calculate the total F475W flux with HST, and double that aperture radius. The greater resolution of HST compared to Subaru, as well as the unknown extent that the ionizing flux from the star clumps penetrates, introduces some uncertainty as to the best aperture size to use. However, the double aperture size encompasses the entire H$\alpha$ peak seen in the Subaru image for Source 6. For Source 16, the aperture size enclosing the entire nearby H$\alpha$ peak is labeled as ``largest aperture'' in the plot, and is about four times the area of the HST aperture.}
	\label{fig:H_alpha}
\end{figure*}

For each source, an estimate of the age was made based on its position in the color-color diagram relative to the tracks, and accounting for error bars on the luminosity of the source. If the source does not fall on the track within the errors, its path was traced to the track using the extinction vector. Uncertainty in age was estimated using the error bars on color. For source 2 and 16, an estimate was difficult as neither error bars, nor the extinction vector can put these source on the track. Large uncertainties in age are given for sources 2 and 16, and an estimate of their actual age was made based on their relative position with respect to the tracks. For source 16, we also compared with estimates of the age and mass based on the H$\alpha$ luminosity of the surrounding H II region (Figure \ref{fig:H_alpha}), discussed below. For each source an estimate of the mass was made based on the age and luminosity of the source, as well as the uncertainties in these parameters. The results of the calculations of mass and age based on the models are shown in Table 1. We find a range of ages between about $1-35$ Myr, and a range in masses of 10$^3$ $\sim$ 10$^5$ M$_{\odot}$.

For some of the star clumps, the masses are so small that the stochasticity of the IMF may begin to have an effect on the colors, affecting our mass and age estimates. In \citet{Calzetti+13}, the author found that a star clump with a mass of 1.7 $\times$ 10$^4$ M$_{\odot}$ was required (such that at least one 30 M$_{\odot}$ star was formed) for a fully sampled Kroupa IMF. For a star cluster of $\sim$ 1 $\times$ 10$^4$ M$_{\odot}$, the effects of stochastic sampling can result in a scatter as large as 20\% in the measured ionizing photon flux; for a star cluster of $\sim$ 1 $\times$ 10$^3$ M$_{\odot}$, scatter can reach 70\% \citep{Calzetti+13, Cervino+02}. This may explain why some sources, such as source 16, with an estimated mass of 6.9$^{+1.6}_{-0.3} $ $\times$10$^{3}$ M$_{\odot}$, fall in regions not predicted by our models.

\begin{table*}
\centering
\caption{Properties of each star clump detected in F275W in the tail of D100. From left to right, in column (1) the source number. In Column (2), (3), and (4), the magnitude and color of the sources.In column (5) the $A_{F475W}$ mag listed is the magnitudes of extinction in the F475W band required for the source to fall on the SB99 tracks. In columns (6) and (7) the age and mass of the sources from SB99. In column (8) the observed half-light radius of the sources in the F475W (discussed in section 3.5.5).}
\label{my-label}
\begin{tabular}{cccccccc}
\hline
\textbf{\#} & \textbf{F475W} & \textbf{F275W-F475W} & \textbf{F475W-F814W} & \textbf{A$_{F475W}$} & \textbf{Age (Myr)} & \textbf{Mass (M$_{\odot}$)} & \textbf{$R_e$ (pc)} \\ \hline
\textbf{1} & 22.2 $\pm$ 0.01 & 0.35 $\pm$ 0.03 & 0.58 $\pm$ 0.01 & 0.3 & 8$^{+5}_{-5}$ & 2.1$^{+1.7}_{-1.4}$ $\times$10$^{5}$ & 104 $\pm$ 21 \\
\textbf{2} & 25.2 $\pm$ 0.04 & -0.29 $\pm$ 0.18 & 1.05 $\pm$ 0.05 & - & 10$^{+5}_{-5}$ & 2.1$^{+1.3}_{-1.2} $ $\times$10$^{4}$ & 75 $\pm$ 24 \\
\textbf{6} & 24.9 $\pm$ 0.02 & -0.009 $\pm$ 0.16 & -0.36 $\pm$ 0.06 & - & 6$^{+1}_{-1}$ & 1.0$^{+0.7}_{-0.1} $ $\times$10$^{4}$ & 62 $\pm$ 20 \\
\textbf{8} & 25.0 $\pm$ 0.03 & 1.04 $\pm$ 0.65 & 1.06 $\pm$ 0.04 & 0.6 & 10$^{+5}_{-5}$ & 2.5$^{+1.6}_{-1.5} $ $\times$10$^{4}$ & 68 $\pm$ 30 \\
\textbf{10} & 24.7 $\pm$ 0.03 & -0.01 $\pm$ 0.22 & 0.71 $\pm$ 0.05 & 0.3 & 10$^{+2}_{-2}$ & 3.3$^{+0.3}_{-1.2} $ $ \times$10$^{4}$ & 74 $\pm$ 31 \\
\textbf{11} & 25.7 $\pm$ 0.04 & 0.21 $\pm$ 0.40 & 1.09 $\pm$ 0.06 & 0.6 & 10$^{+5}_{-5}$ & 1.3$^{+0.9}_{-0.8} $ $\times$10$^{4}$ & 47 $\pm$ 25 \\
\textbf{16} & 25.1 $\pm$ 0.04 & -0.51 $\pm$ 0.19 & -0.30 $\pm$ 0.15 & 0.1 & 1$^{+5}_{-0.9}$ & 6.9$^{+1.6}_{-0.3} $ $\times$10$^{3}$ & 51 $\pm$ 23 \\
\textbf{27} & 26.9 $\pm$ 0.10 & -0.22 $\pm$ 0.56 & 0.15 $\pm$ 0.24 & - & 9$^{+15}_{-2}$ & 4.1$^{+5.7}_{-2.4} $ $\times$10$^{3}$ & $\le$ 36 \\
\textbf{28} & 25.9 $\pm$ 0.06 & 0.43 $\pm$ 0.57 & 0.20 $\pm$ 0.14 & 0.1 & 35$^{+20}_{-20}$ & 3.1$^{+0.5}_{-1.2} $ $\times$10$^{4}$ & 63 $\pm$ 21 \\
\textbf{31} & 25.3 $\pm$ 0.04 & 1.05 $\pm$ 0.61 & 0.74 $\pm$ 0.05 & 0.5 & 9$^{+16}_{-2}$ & 1.8$^{+2.6}_{-1.0} $ $\times$10$^{4}$ & 51 $\pm$ 23
\end{tabular}
\end{table*}

The inclusion of Starburst99 models with nebular lines allows us to compare the measured H$\alpha$ luminosity from the ground-based Subaru telescope observations with those predicted for the star clumps with possible H II regions. These would be sources with ages younger than 10 Myr. Four sources from Table 1 fit this criteria, but source 31 has a large uncertainty in the age, and does not appear to have any nearby H$\alpha$ peak suggestive of an H II region (although at an age of 9 Myr, the H$\alpha$ flux of the H II region approaches the detection limit of the Subaru observations). Thus we compare just sources 16, 6, and 1, to our model, the youngest sources with ages of 1$^{+5}_{-0.9}$, 6$^{+1}_{-1}$, and 8$^{+5}_{-5}$, Myr respectively.

Sources 1 \& 6 show excellent agreement with the predicted H$\alpha$ flux, adding an independent confirmation of our estimates of the age and mass of these sources. For Source 16, there is a large uncertainty in the age, and correct aperture size. The source agrees within the error bars, but shows a generally lower than expected H$\alpha$ flux. This may be a factor of the stochasticity of the IMF for such a low mass source, contributing to a different relation between stellar mass and H$\alpha$ flux than predicted from the IMF of our model. Dust extinction in the H II region surrounding this source is also unknown. However, the lower than expected H$\alpha$ luminosity may also be a sign of RPS stripping of the H II region itself. If enough gas is stripped, some Ly-$\alpha$ photons escape the H II region, and the source will produce less H$\alpha$ flux from recombination than predicted. This effect could cause the SFR based on H$\alpha$ luminosity of RPS tails to be underestimated (Kenney, J.D.P., et al. \textit{in prep}).

Along with an instantaneous burst model, we also experimented with a model in which sources formed within a duration 10 Myr burst (with a boxcar shape). However, we found that about half the sources shown in Figure \ref{fig:tracks_labeled} (those with F475W-F814W $\gtrsim$ 0.8) no longer fell near the tracks, since the extended rightward jog in the track at ages $\sim 10-30$ Myr in the color-color diagram of Figure \ref{fig:tracks_labeled} largely disappears. This suggests that the instantaneous burst model is preferable to the 10 Myr burst model. For sources that fell nearer the 10 Myr burst track, the best fitting average age is older, and the best fitting mass is larger by about a factor of two. This was because sources composed of older stellar populations with the same luminosity as sources composed of younger stellar populations tend to be more massive. However, our comparison of the predicted and observed H$\alpha$ luminosities of some sources, shown in Figure \ref{fig:H_alpha}, supports the age and mass estimates for the sources from our instantaneous models. Thus, we believe an instantaneous burst model is the best approximation for the star formation history of these sources.

\subsubsection{Comparison of star clump properties in D100 with other RPS tails}

Our results, along with other studies of ram pressure stripped \citep{Cortese+07, Yagi+13} and tidal \citep{Boselli+18} tails, find ages of stars in tails to be $\lesssim 100$ Myr, and with masses between $10^{3}-10^{6}$ M$_{\odot}$ with masses on the higher end of this range being rarer. While the highest masses are similar to those of globular clusters, we find in this study that the tail star clumps are probably not bound (see Section 3.5.5), so these sources are probably not single star clusters. The consistent ages of $\lesssim 100$ Myr are likely due to observational constraints; it is much easier to observe younger stars, as luminosity declines rapidly with age. One noteworthy case of older stars found in a ram pressure stripped tail is in the dwarf galaxy IC 3418 \citep{Fumagalli+11}, in which the ages of some stars in the tail, after removal of contaminating background sources \citep{Kenney+14}, were found to be as old as 300 Myr. It is challenging but important to detect older stars in tails, in order to piece together their evolutionary histories.

\subsubsection{The SFR of the tail}

We calculate the SFR of the tail from our estimates of age and mass from SB99 models. The most dominant source in stellar mass is source 1 at the base of the tail ($\sim$ 3.5'' from the nucleus), with a mass about equal to that of all other sources combined. We choose to exclude this region because we wish to ultimately calculate the star formation efficiency of the main part of the tail, and this source is not resolved from the disk of D100 in the IRAM $\sim$ 30'' aperture observations of D100 from \citet{Jachym+17}.

\begin{figure*}
	\plotone{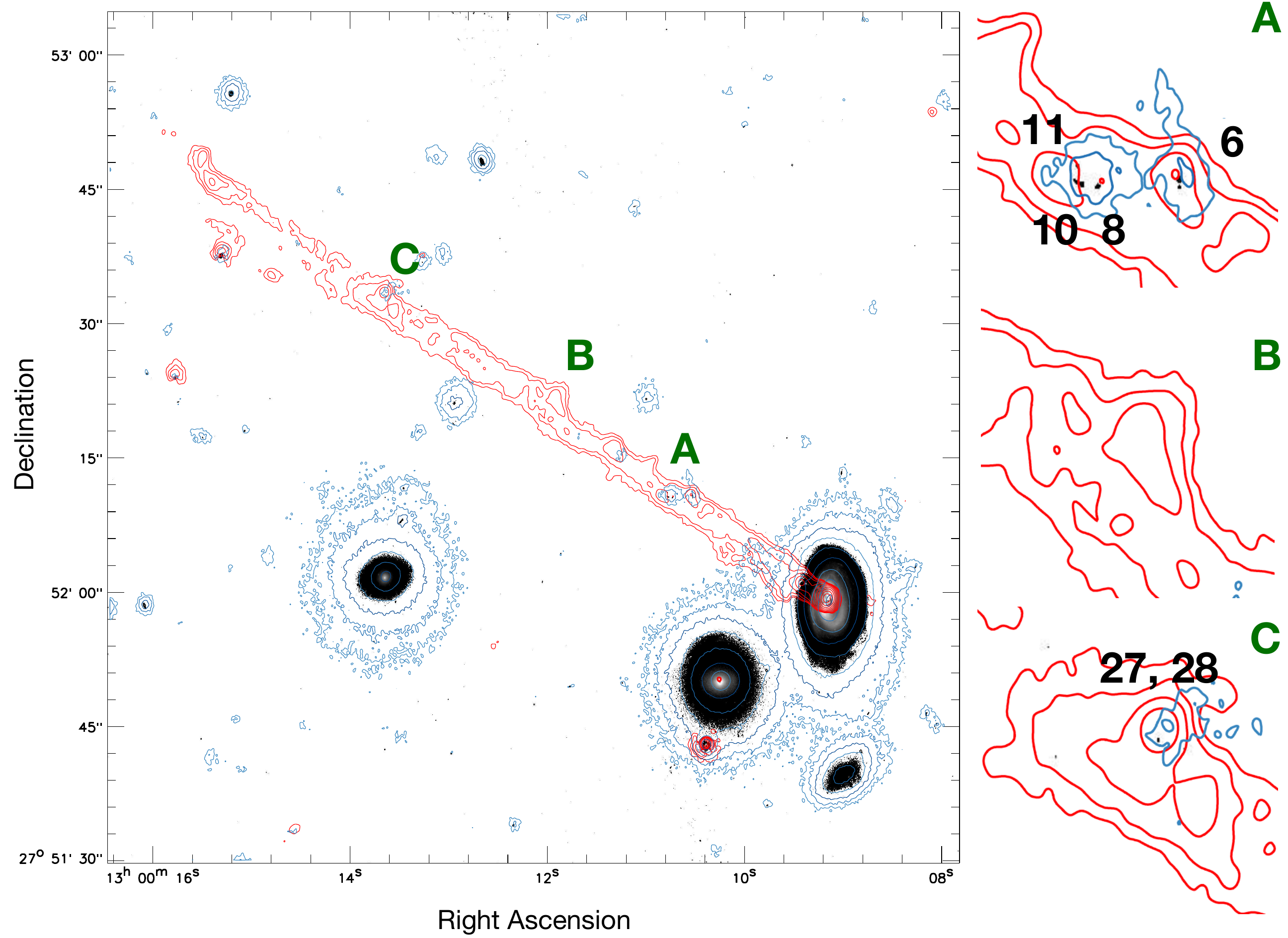}
	\caption{\textbf{Greyscale}: HST F475W, \textbf{Red}: Subaru H$\alpha$, \textbf{Blue}: $u$-band CFHT observations from \citet{Smith+10}. Three zoom-ins are chosen, both \textbf{A} and \textbf{C} have associated stars identified in both HST and $u$-band, and are coincident with nearby H$\alpha$ peaks, while \textbf{B} is an example of a region with a strong H$\alpha$ peak, and no associated stars detected in either HST or $u$-band. $u$-band contour vary from 29.9 to 23.12 mag/arcsec$^2$ by increments of 0.9 mag/arcsec$^2$, and H$\alpha$ contours in the tail vary from 4.8 to 160,000 $\times \, 10^{-18}$ erg s$^{-1}$ cm$^{-2}$ arcsec$^{-2}$ by a factor of two for each contour level.}
	\label{fig:UV-HST}
\end{figure*}

The total stellar mass of all the other sources detected by HST in the tail is 1.6$^{+0.8}_{-0.7}$ $\times \,10^5$ M$_{\odot}$. However, we must consider the fact that the limiting magnitude of our F275W observations means that we are not sensitive to young star clumps below a certain mass that may be present. We use the CFHT $u$-band data from \citet{Smith+10}, which is more sensitive to diffuse stellar emission, to estimate the mass of stars below our limit. One concern about this data is that there could be line contamination from the 3727\AA \, [OII] line within the $u$-band. However, we find that line contamination is unlikely to be a large effect. In Figure \ref{fig:UV-HST}, we show that the $u$-band data aligns spatially with the HST observed clumps, while the H$\alpha$ is slightly offset, or in some cases unassociated with the $u$-band, seen in the Figure \ref{fig:UV-HST}. Thus, it seems likely that the $u$-band flux is dominated by stellar continuum emission.

This displacement in the local peak of the H$\alpha$ and the location of some of the stellar sources is also quite interesting in its own right. One would expect sources of 10 Myr or less in age to have associated H II regions. In all but source 31 (which has significant uncertainty in the age) this appears to be true. However, these peaks can be seen, to varying degree, slightly offset downstream from the stellar sources, in both Figure \ref{fig:UV-HST}, and the cutouts in Figure \ref{fig:stamps}. The downstream displacement in H$\alpha$ could be due to the original gas in the H II region in which the stars were formed being stripped further downstream by the RPS. This phenomenon has also been seen in other ram pressure stripped galaxies, such as RB 199 \citep{Yoshida+12}, and in the tail of NGC 4388 \citep{Yagi+13}.

Given that $\sim$ 72\% of the $u$-band flux is found in the clumps of stars we identify with HST, and the other 28\% is found in diffuse emission throughout the tail, we assume 28\% the stars by mass in the tail are located outside our detected clumps, and below the detection limits of our HST observations. Thus, there is a total mass of young stars in the tail of 2.1 $^{+1.0}_{-0.9}$ $\times \,10^5$ M$_{\odot}$. This is assuming that the ages and colors of these stars undetected by HST are, on average, similar to those we have detected.

The estimate of the timescale of star formation to which we are sensitive is based on two factors. The first is the oldest star cluster we detect, source 28 with an age of $35 \pm \, 20$ Myr. The second factor is the detection limit in the F275W of our data; theoretically, we should be able to detect a 10$^4$ M$_{\odot}$ star clump up to 50 Myr in age, after accounting for extinction. We use $35 \pm \, 20$ Myr as the maximum age of star formation to which we are sensitive. With this stellar mass and timescale, we estimate the SFR of the tail as:

%Calculations based on our stellar mass tracks from Figure \ref{fig:colormag1} have determined that a star clump of $10^4$ M$_{\odot}$, a typical mass based on our source sample, would be detectable to a maximum age of $\sim$ 75 Myr

$$\mathrm{SFR_{tail}}=\frac{2.1 \times 10^5 \, M_{\odot}}{3.5 \times 10^7 \, \mathrm{yr}}=6.0^{+3}_{-3} \times 10^{-3} \, M_{\odot} \, \mathrm{yr}^{-1}.$$

Our calculations on the SFR of the tail show that H$\alpha$ flux should not be used as a proxy for star formation in ram pressure stripped tails. The total H$\alpha$ flux is calculated from the ground-based Subaru observations \citep{Yagi+10}, which, as described previously in Section 2, have been corrected for the [NII] and [SII] lines, as well as over-subtraction in the $R$-band. The total flux of the tail excluding the area around source 1 is approximately 6.5$\times 10^{-15}$ ergs/s/cm$^2$. This is after applying a correction for the estimated dust extinction in the tail, based on the average extinction for the star clumps listed in Table 1, resulting in approximately 0.28 magnitudes of extinction at H$\alpha$. At the distance of D100, this corresponds to a total luminosity of $L_{H\alpha} = 7.78 \times 10^{39}$ ergs/s, which corresponds to a SFR of 0.042 $M_{\odot}$ yr$^{-1}$, using the coefficients for translating H$\alpha$ flux to SFR from \citet{Kennicutt+09}, assuming a Kroupa IMF. However, our HST analysis has found a measured SFR of only $0.006^{+0.003}_{-0.003}$ $M_{\odot}$ yr$^{-1}$, a factor of seven less. This indicates that most of the H$\alpha$ emission in the tail must be the product of some mechanism other than star formation. 

It should be noted that for star clumps that do not fully stochastically sample the IMF, the H$\alpha$/UV ratio has been shown to be lower than fully sampled clusters \citep{Fumagalli+11}. Furthermore, short term variations in the star formation history, such as may occur in the extreme environment of a ram pressure stripped tail, have also been shown to lower this ratio, when compared to normal star formation histories \citep{Boselli+09, Emami+18}. Thus there may be additional uncertainty to consider when comparing the expected SFR of the tail with a standard formula such as that by \citet{Kennicutt+09}. However, these factors alone would not be enough on their own to explain a factor of seven difference between the predicted and measured H$\alpha$ luminosity. Furthermore, much of the H$\alpha$ emission in the tail is extended and smooth, and not patchy as would be expected from H$\alpha$ emission associated with star formation.

Our findings on the low measured SFR in comparison to the prediction from H$\alpha$ is in contrast to other recent studies of ram pressure stripped tails. A study of a ram pressure stripped `jellyfish' galaxy in the GASP sample by \citet{George+18} finds good concordance between the measured SFR from the UV, and that predicted from the H$\alpha$ emission in the tail. Other studies of GASP galaxies, such as \citet{Poggianti+19}, have also found that the dominant ionization mechanism for H$\alpha$ excitation in these tails is photoionization from young stars. This is quite different than our results for the tail of D100, in which we find levels of star formation seven times lower than those predicted from H$\alpha$, suggesting a dominant mechanism, or mechanisms, other than photoionization from young stars is present in the tail. Thus, one would not want to make a general conclusion about RPS galaxies from only a subset of the population.

\subsubsection{The star formation efficiency of the tail}

To estimate the amount of molecular gas in the tail of D100, we use the number quoted in \citet{Jachym+17}, M$_{\mathrm{H}2}$ $\sim$ 10$^9 \, M_{\odot}$. We also have an upper limit on the H I in the tail of 0.5 $\times \, 10^8 \, M_{\odot}$ \citep{Jachym+17}. The mass of H$_2$ and H I in the tail allows us to calculate the gas consumption timescale, ``star formation efficiency", of the tail as:

$$\mathrm{SFE_{tail}}=\frac{6.0^{+3}_{-3} \times 10^{-3}  \, M_{\odot} \mathrm{yr}^{-1}}{10^9 \, M_{\odot}} = 6.0^{+3}_{-3} \times10^{-12} \, \mathrm{yr}^{-1}.$$

\begin{figure*}
	\plotone{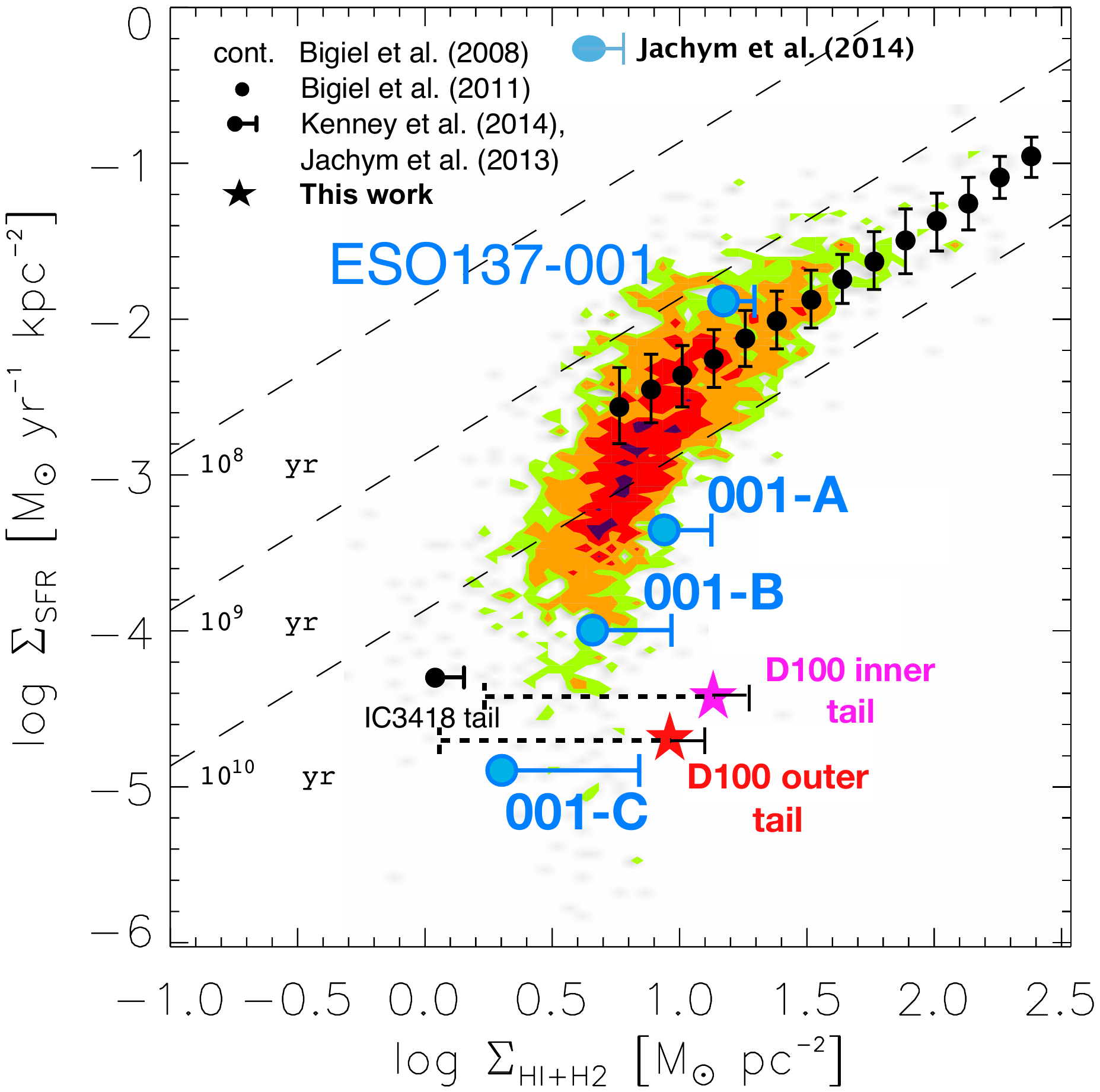}
	\caption{SFR surface density versus gas surface density, from \citet{Jachym+14}, with our data from D100 added. The lower error bars show what the results would be with a factor of 10 difference in the CO$-$H$_2$ relation. The upper error bar on the gas mass is given from the limit on the mass of H I in the tail. In blue points are the body, and three tail regions of ESO 137-001, with point A being closest to the body of the galaxy, and C being furthest away. Filled circles show molecular gas, and the bars show upper limits on HI. Contours show SFRs and efficiencies sampled from the disks of seven nearby spirals from \citet{Bigiel+08}. The black dots show average molecular gas depletion time measured in 30 nearby galaxies. For comparison, the tail of the Virgo cluster dwarf galaxy IC3418 is also included. The blue pentagons show data from \citet{Moretti+18} from four ram pressure stripped tails in the GASP sample. They represent lower limits on both the star formation and gas surface density (see \citet{Moretti+18} for details), thus the SFE in these tails could be even higher than shown here.}
	\label{fig:Pavel_14_tail_add}
\end{figure*}

The overall star formation efficiency of the tail is $\sim$ $6 \, \times$ 10$^{-12}$ yr$^{-1}$, quite low, corresponding to a gas depletion timescale of 1.7 $\times$ 10$^{11}$ years. This is nearly 2 orders of magnitude lower than the inner disks of normal spirals \citet{Bigiel+08}, but comparable to the SFE of the outer disks of spirals and a couple other RPS tails. However, if our estimate of the gas mass is accurate, it is the highest gas surface density at which such a low SFE has been measured. We compare this estimate of star formation efficiency to that in the disks of nearby spirals, as well as two other ram pressure stripped tails in Figure \ref{fig:Pavel_14_tail_add}. 

We also calculate the surface density of gas in the tail. We assume that the molecular gas detected in the apertures is all concentrated in the region of the tail visible in H$\alpha$. Given a tail length of 60 kpc and a width of 1.5 kpc, the total area of the tail is 90 kpc$^2$. This gives an average gas surface density over the tail of $\Sigma_{gas}= 11 \, M_{\odot} \, \mathrm{pc}^{-2}$, and a SFR density of $\Sigma_{SFR} =6.6^{+3.3}_{-3.3} \times10^{-5} \, M_{\odot}  \,\mathrm{yr}^{-1} \, \mathrm{kpc}^{-2}$.

In order to see whether there is a gradient in SFE based on radial distance from the body of the host galaxy, we divide the tail into two parts, the inner tail, extending along the first 30 kpc of the tail, and the outer tail extending along the last 30 kpc. Based on the fraction of molecular gas per aperture along the tail, we estimate that 57\% of the total flux from molecular gas is detected in the inner tail, while 43\% is found in the outer tail. Given a total mass of molecular gas in the tail of $\sim10^9$ M$_{\odot}$, this results in the surface density of gas in the inner tail being 13 $M_{\odot} \, \mathrm{pc}^{-2}$, and in outer tail, 10 $M_{\odot} \, \mathrm{pc}^{-2}$. Furthermore, only sources 27, 28, and 31 are found in the outer tail. The resulting surface densities of gas and star formation are plotted in Figure \ref{fig:Pavel_14_tail_add}.

Overall, D100 exhibits star formation efficiency comparable to, and possibly even lower than that found in the tail of ESO 137-001. However, we must consider that the amount of molecular gas relies on an accurate H$_2$-CO conversion factor. \citet{Jachym+17} point out that the conversion factor may vary widely throughout the tail, especially from empirical values measured in the disks of galaxies due to the different conditions in this environment. We show an error bars on the mass of molecular gas in the tail of a factor of 10 less in Figure \ref{fig:Pavel_14_tail_add} to account for this. Even accounting for such an extreme variance in the H$_2$-CO conversion factor, the SFE in the tail of D100 is much lower than that of the inner disks of spirals from the \citet{Bigiel+08} sample. The upper error bar on the gas mass is given from the limit on the mass of H I in the tail.

Similar to what is seen in ESO 137-001, we also see the same trend in star formation efficiency falling with distance from the body of the galaxy, with the outer tail being about two times less efficient than the inner tail. This suggests that this may be the case overall for ram pressure stripped galaxies. This could be explained by the idea that the less dense gas will be accelerated by ram pressure more easily than the densest gas, meaning that more dense gas, i.e. the sites of star formation, will be located preferentially in the inner tail \citep{Jachym+14, Jachym+17}. A key difference between D100 and ESO 137-001 is that the gas surface density in the tail of D100 is almost an order of magnitude greater than that of ESO 137-001. While ESO 137-001 has similar SFR and gas surface density as the outer disks of nearby spirals, the tail of D100 has a gas surface density of $\sim$ 9-12 $M_{\odot} \, \mathrm{pc}^{-2}$. This corresponds to the area of disks where the star formation surface density in normal spiral disks is almost two orders of magnitude greater.

We note that the beam size for the CO(1-0) observations are not the same in the three plotted galaxies. The ratio in beam area between ESO 137-001 and D100 is $\sim$ 2.5, so ESO 137-001 is more beam diluted than D100, which could result in it having a lower beam-averaged gas surface density (and SFR/area), by a factor of $\sim$ 2. This factor would not change our overall conclusions regarding the star formation efficiency comparison between the galaxies.

\subsubsection{Are young tail stars in star clusters?}

\begin{figure*}
	\plotone{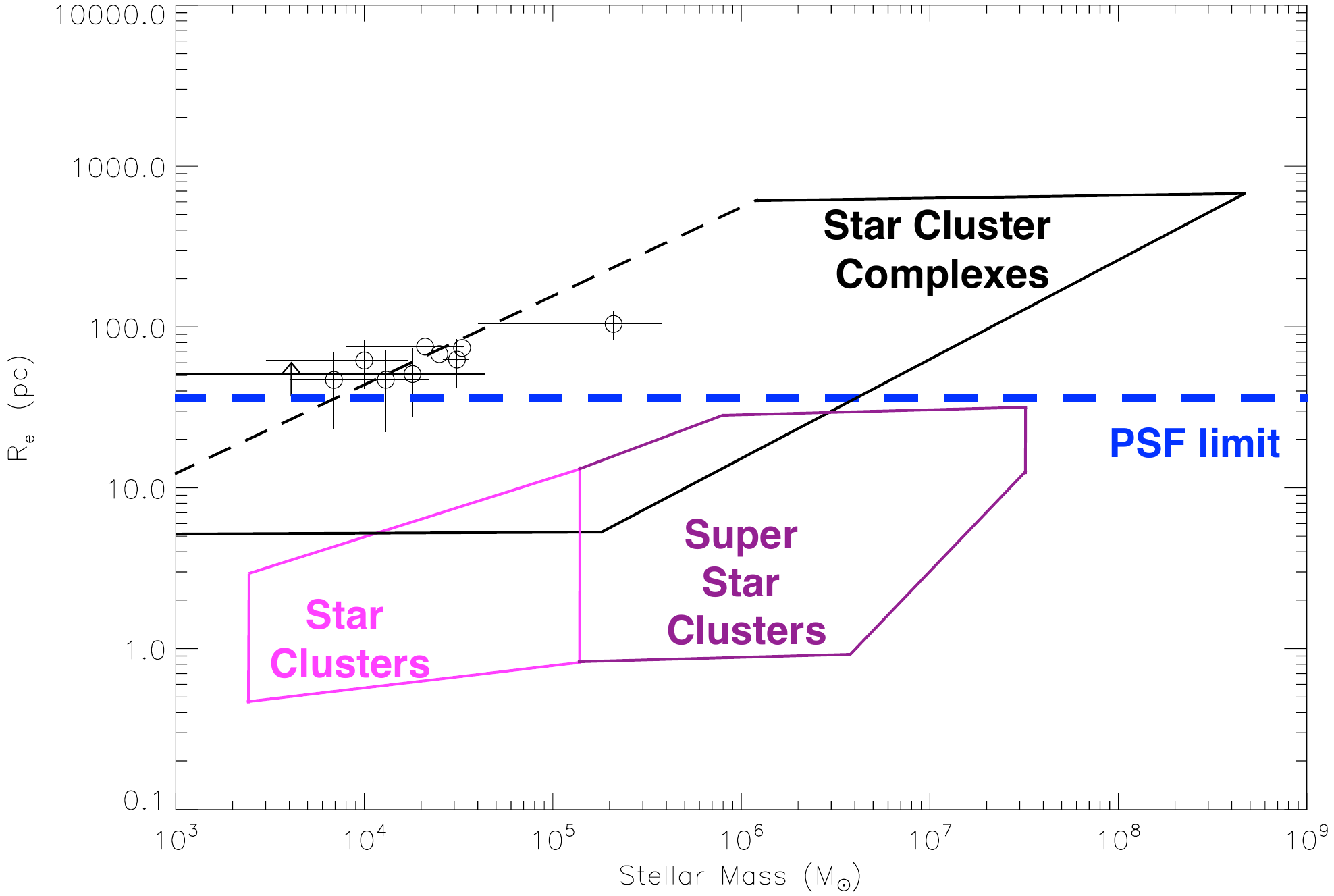}
	\caption{The effective radii of the star clusters found within the tail, calculated by taking the geometric mean of the effective semi-minor and semi-major axes, are plotted versus the stellar masses. The polygons of different colors show the general parameter space of star clusters, super stellar clusters, and star cluster complexes from an empirical sample compiled by \citet{Bouwens+17}. Star clusters and super stellar clusters are taken from a sample in the local universe (z $\sim$ 0), while star cluster complex sizes and masses come from a large sample of complexes in galaxies from z $=0-3$. The dashed black line indicated the limits of the sensitivity of the sample from \citet{Bouwens+17}. Our deep observations on the relatively nearby Coma cluster allows us to sample a fainter cluster population than \citet{Bouwens+17}. The blue line shows the PSF limit of the size of a resolved star clump for the HST observations, 36.5 pc, and the upper limit sign is for source 27.}
	\label{fig:clump_sizes}
\end{figure*}

With the resolution of HST we get useful information on the sizes of the tail star clumps. In Figure \ref{fig:clump_sizes} we plot the effective (half-light) radii R$_{e}$ of the stellar sources versus their stellar masses, and compare to the effective radii and masses of known star clusters, super star clusters, and star cluster complexes from \citet{Bouwens+17}. The source sizes have been corrected for the PSF of the HST F475W filter ($\sim$ 0.08'' for our data). Only one source, source 27, the least massive star clump, is unresolved and thus is labeled as an upper limit in the plot. The sizes of the sources in D100 are much larger than single star clusters, which are gravitationally bound, and are consistent with these sources being large, unbound star cluster complexes. This suggests that the star cluster complexes we are analyzing are gravitationally unbound, and will disperse over time. This would make it difficult to detect the older stars formed in ram pressure stripped tails, and to determine the total contribution to intracluster light from ram pressure stripped tails.

\section{On the morphology of the tail}

An especially remarkable feature of D100 is its simple, long, straight, and relatively unbroadening H$\alpha$ tail \citep{Yagi+07}. We gain new insights into the physical processes that shape the tail from our HST imaging data, both from the dust morphology viewed in the base of the tail, and from the temporal progression of outside-in disk stripping that we have measured from HST colors, that relates directly to the broadening of the tail. 

The color HST image in Figure \ref{fig:colorstreams} shows coherent, filamentary dust structures, especially visible at both edges of the tail. The northernmost filament shows a straight, continuous filament extending a total of 4'' (1.9 kpc), suggesting little turbulence in the flow of the gas. The long, straight, and relatively unbroadening H$\alpha$ tail (Figure \ref{fig:sidebyside}) also suggests a minimum of turbulence. Due to how relatively simple and well-defined the tail of D100 is, it is easier to study than tails with many complex components from different radii. It is an excellent galaxy to compare to simulations, to investigate the factors that contribute to the overall width and structure of tails.

Other observed ram pressure stripped tails, such as the well studied tail of ESO 137-001, tend to be messier, i.e. with discrete components at a range of distances from the tail center, and overall more broad than the tail of D100. In Figure 2 of \citet{Jachym+14}, along with the main tail, one other filamentary gas tail at larger galactic radii can be seen in ESO 137-001. This would be due to the lower density ISM in the inner radii of the galaxy being stripped before the outer disk completely loses its gas. This is likely the case for low inclination (closer to face-on) stripping when diffuse gas is stripped from a region before denser ISM, leaving pockets in the gas disk where hydrodynamic instabilities can form, that increase the stripping rate \citep{Quilis+00}. Other galaxies in Coma with ram pressure stripped tails such as IC4040 \citep{Yoshida+12} and RB199 \citep{Yoshida+08} also show these characteristic multi-component tails likely due to stripping at multiple radii at once. D100 may have a simpler, single-component tail structure, with  inhibited stripping at multiple radii, due to undergoing a highly inclined stripping event (see Figure 7 of \citet{Jachym+17} for a visualization). In highly-inclined stripping cases there is both a larger path length for gas to travel through the disk, and a higher projected gas surface density.  These factors make it difficult for ram pressure to punch holes through the disk.

Another remarkable feature of the D100 tail is its narrow, relatively unchanging width over its length. Our observations have shown that the dust tail expands from an initial diameter of 0.95 kpc to, a diameter of 1.4 kpc at the edge of the disk of the galaxy, an increase of $\sim$ 50\%. The H$\alpha$ tail extends 60 kpc from the center of the galaxy, and broadens slightly from a half-width of 0.95 kpc at the center to 1.7 kpc at the end of the tail.  There are several plausible mechanisms by which tails could broaden, and the fact that so little broadening is seen means that their impact on the tail of D100 is minimal. We briefly discuss each possible mechanism: radial progression of outside-in stripping, angular momentum from disk rotation, gas pressure, and turbulence (including influence from magnetic fields).

In the absence of any other broadening mechanism, the tail should broaden simply due to the radial progression of outside-in disk stripping. From the HST colors, we have determined (in Section 3.4.1) the star formation quenching time as a function of radius, and found that the stripping radius has progressed from $r \sim 1.3-2.3$ kpc to $r \sim 0.25$ kpc in the last 280 Myr.  This matches very well the measured $\sim$ 2 kpc half-width of the outer tail and the $\sim$ 250 Myr age of the outer tail estimated from gas kinematics from \citet{Jachym+17}. Thus, nearly all of the tail broadening is consistent with the radial progression of outside-in disk stripping, putting a strong limit on all the other factors which can cause broadening. Furthermore, this suggests gas stripped from further out in the disk of D100 should be found even further downstream. This gas may be part of a broader, fainter (not currently detected), and older tail component.

Another factor in the lack of tail broadening could be due to the near edge-on stripping of D100. The momentum of the gas in the disk carries some orbital velocity, related to the distance from galaxy center and the mass of the host galaxy. As this gas is stripped, it will expand outward since it is no longer forced to rotate about the galaxy center. \citet{Roediger+05} showed in simulations that azimuthal asymmetries in the radius of origin of gas tails in RPS galaxies resulted for inclinations less than 30 degrees (like D100) due to the angular momentum of gas in the disk. In nearly face-on stripping, the angle of incidence of RPS is near perpendicular to the angle of rotation of gas in the disk, likely leading to neither inhibition, nor promotion, of broadening. In the nearly edge-on stripping case, however, during part of its orbit, the gas will rotate at a near parallel or anti-parallel angle to the incidence of RPS. The momentum change from ram pressure would thus sometimes act opposite the angle of rotation, reducing the angular momentum of the gas as it is stripped. This could result in less broadening, as opposed to other observed tails in galaxies that are stripped closer to face-on. It should be noted D100 is a relatively low-mass spiral, with the rotation speed estimated to be only about 50 km s$^{-1}$ in the inner galaxy \citep{Caldwell+99}. Such low angular velocity in the innermost galaxy could also result in little to no broadening.

Another explanation for the lack of tail broadening comes from simulations that focus on modeling gas cooling, such as that by \citet{Tonnesen+10}, which show that narrower tails result from including gas cooling in their simulation. This is due to the reduced pressure of the gas in the tail with respect to the intercluster medium (ICM). If the ISM of the tail is overpressured with respect to ICM it will expand. The tail from their paper that has cooled to $\sim$8,000 K, similar to the temperature of the gas seen in the H$\alpha$ image, appears similar to the H$\alpha$ tail of D100 in its lack of significant broadening.

Finally, the turbulence of both gas in the tail, and in the surrounding ICM, is a factor that could influence broadening. Work by \citet{Roediger+08b} suggests that flaring of the gas tail is determined by the turbulence in the ICM flow past the galaxy. By varying the viscosity in their hydrodynamical simulation, \citet{Roediger+08a} found that an inviscid ICM leads to more turbulence and vortices in the ram pressure stripped tail of the simulated galaxy. Consequently, a high viscosity in the ICM in the area of D100 would lead to the suppression of hydrodynamical instabilities, and result in a more straight, and unbroadening tail. That said, studies of ram pressure stripping of elliptical galaxies in the Virgo cluster found no need to invoke any additional factor for ICM viscosity to explain the observed tails  \citep{Roediger+15a, Roediger+15b, Kraft+17}. Coma, however, is significantly hotter than Virgo, so we would expect the influence from viscosity to be much higher, due to its strong temperature dependence. D100 would be an excellent place to further study ICM viscosity and thermal conductivity.

The influence of magnetic fields could also reduce turbulence in the tail. \citet{Ruszkowski+14} found that, in a comparison between an MHD simulation (with fields only in the ICM, not the body of the galaxy), and a hydrodynamical simulation, the morphology of RPS tails changed significantly. In MHD simulations, magnetic fields inhibit thermal conduction, and result in the formation of long, filamentary structures in the gas tail. The tail that formed in the MHD simulation was also much narrower than the corresponding, purely hydrodynamical case, in which a much clumpier, disparate, tail resulted. Such structure has been noted in the H$\alpha$ tail of NGC 4569 in Virgo by \citet{Boselli+16}, who also cited it as evidence of the influence of magnetic fields in the tail. Our even higher resolution view of the dust in D100 with HST reveals kpc scale filaments of dust, not previously seen in the dust in any other RPS tail. Magnetic fields permeating the stripped gas may play a role in inhibiting turbulence, leading to a narrower tail.  

The modest amount of broadening in the H$\alpha$ tail is consistent with being caused by the radial progression of outside-in stripping over the last $\sim$ 250 Myr that we have measured. The other factors that could cause broadening, specifically angular momentum from disk rotation, gas pressure, and turbulence, are apparently not significant in the tail of D100. More work is needed to understand the influence of magnetic fields in RPS galaxies; the importance of magnetic fields in the formation of ISM substructure has been noted in other RPS influenced galaxies \citep{Kenney+15}.

\section{Summary}

We have presented new HST F275W, F475W, and F814W observations of the galaxy D100 in the Coma cluster, known for its spectacular 60 $\times$ 1.5 kpc ram pressure stripped H$\alpha$ tail. This tail was previously found to host gas hot enough to emit in the soft X-ray, as well as containing $\sim$ 10$^9$ M$_{\odot}$ of cold molecular gas. Given the amount of molecular gas, star formation in the tail was expected, but not measured quantitatively. Our new data has allowed us to characterize the amount and efficiency of star formation, as well as the ages, masses, and sizes of star complexes that formed in that tail. We also analyzed the star formation histories of the main bodies of D100 and its two close companions, and shown that their quenching histories suggest evidence of ram pressure stripping in all three.

\begin{description}
	\item[$\bullet$ Star formation in the tail] Through analysis of the colors and magnitudes from HST, we have constrained the ages and masses of the tail star clumps from comparison with the Starburst99 SSP model. We find a SFR in the tail of 6.0 $\times \,10^{-3}$ M$_{\odot}$ yr$^{-1}$ over the last 35 Myr, and a star formation efficiency of $ 6 \times \,10^{-12}$ yr$^{-1}$. Overall, the star formation efficiency is a factor of $\sim$ 2 times higher in the half of the tail closer to the galaxy. Furthermore, the SFR is a factor of 7 times less than would be predicted by the H$\alpha$ flux of the tail, demonstrating some other excitation mechanism is dominant in the H$\alpha$ tail. This in contrast to some recently published results from the GASP group, which found good agreement between the H$\alpha$ flux and measured SFR \citep{George+18, Poggianti+19}.
	
	\item[$\bullet$ Star clump sizes] We have found from analysis of the stellar masses and sizes of the star complexes in the tail of D100, and comparison to a large sample of observed star clusters and complexes, that the tail star clumps are likely to be gravitationally unbound complexes. They have masses of $\sim$ $10^3-10^5$ M$_{\odot}$, and sizes (based on $R_e$) of $50-100$ pc, much larger than any known bound star clusters. Thus, they should disperse over time, rendering older stellar clumps formed in the tail more diffuse, and harder to detect.
	
	\item[$\bullet$ Outside-in quenching] From color analyses, we find radial age gradients due to outside-in quenching in D100 and its apparent neighbor D99. For a model star formation history with a 2\% burst at the time of quenching, D100 has a quenching time of $\sim$ 280 Myr at $r=2$ kpc and $\le$ 50 Myr at $r=0.5$ kpc, and an ongoing starburst inside $r=0.5$ kpc. D99 is completely quenched, with a quenching time of ~1 Gyr at $r=3$ kpc, and $\sim$ 300 Myr in the center.
	
	\item[$\bullet$ Longevity of spiral arms] While the disk of D100 outside the central $r \sim 500$ pc has been completely stripped of gas and ceased star formation $100-400$ Myr ago, strong spiral structure is still visible. In D100's apparent neighbor D99, which is similar in mass and basic structure to D100 but has an outer disk quenching age of $\sim$ 1 Gyr, faint spiral structure is still visible. Unsharp masking reveals two main spiral arms with amplitudes of only $\sim$ 1.5\%. This suggests the time for a stripped spiral to evolve into an S0 type galaxy is on the order of $\sim$ 1 Gyr.
	
	\item[$\bullet$ Substructure in the tail and little broadening] HST images reveal parallel kiloparsec-length dust filaments not previously seen in any other RPS tail. Such filaments are seen in MHD simulations, but not purely hydrodynamical simulations, suggesting that magnetic fields are important in RPS tails. Magnetic fields permeating the stripped gas may play a role in inhibiting turbulence, leading to a narrower tail.  The modest amount of broadening in the H$\alpha$ tail is consistent with being caused by the radial progression of outside-in stripping over the last $\sim$ 250 Myr that we have measured. The other factors that could cause broadening, specifically angular momentum from disk rotation, gas pressure, and turbulence, are apparently not significant in the tail of D100.
	
\end{description}

\acknowledgments

We thank the referee for their helpful comments. W. Cramer, J. Kenney, \& M. Sun acknowledge the support of STScI Grant HST-GO-14361.003. Based on observations made with the NASA/ESA Hubble Space Telescope, obtained June-July of 2016, at the Space Telescope Science Institute, which is operated by the Association of Universities for Research in Astronomy, Inc., under NASA contract NAS 5-26555. These observations are associated with program 14361 (PI:Sun), and additional archival data from program 14182 (PI:Puzia). M. Sun would like to acknowledge support from NSF grant 1714764, M. Sun would also like to acknowledge support from Chandra Award GO6-17111X. This work is based in part on data collected by the Subaru Telescope, which is operated by the National Astronomical Observatory of Japan. We also thank Russell Smith for facilitating, and Stephen Gwynn for providing the reduced CFHT data from MegaPipe. P.J. acknowledges support by the project LM2015067 of the Ministry of Education, Youth and Sports of the Czech Republic. Funding for the Sloan Digital Sky Survey IV has been provided by the Alfred P. Sloan Foundation, the U.S. Department of Energy Office of Science, and the Participating Institutions. SDSS acknowledges support and resources from the Center for High-Performance Computing at the University of Utah. The SDSS web site is www.sdss.org. This research has made use of the SVO Filter Profile Service (http://svo2.cab.inta-csic.es/theory/fps/) supported from the Spanish MINECO through grant AyA2014-55216. Fantastic false color images of the reduced HST data were provided by the STScI imaging team led by J. DePasquale.

\clearpage

\bibliographystyle{apalike}
\bibliography{Bibliography}

\end{document}